\documentclass[aip, rsi, reprint, amsmath, amssymb, graphicx, floatfix]{revtex4-1}
\usepackage{amsmath,amsfonts,amssymb}
\usepackage{graphicx}
\usepackage[colorlinks=true, allcolors=blue]{hyperref}
\usepackage{dcolumn}
\usepackage{bm}
\usepackage[utf8]{inputenc}
\usepackage[T1]{fontenc}
\usepackage{mathptmx}
\usepackage{etoolbox}
\usepackage{multirow}
\usepackage{enumitem}
\usepackage{makecell}


\newcommand{\micron}[0]{\textmu m}
\begin{document}

\title{Diamond-loaded polyimide aerogel scattering filters and their applications in astrophysical and planetary science observations}
\author{Kyle R. Helson}
 \affiliation{The University of Maryland, Baltimore County. Baltimore, Maryland, USA}
  \email{kyle.helson@nasa.gov}
 \affiliation{NASA Goddard Space Flight Center, Greenbelt, Maryland, USA}

\author{Carol Yan Yan Chan}
    \affiliation{The Johns Hopkins University, Baltimore, Maryland, USA}
\author{Stefan Arseneau}
\affiliation{Boston University, Boston, Massachusetts, USA}
\author{Alyssa Barlis}
\affiliation{NASA Goddard Space Flight Center, Greenbelt, Maryland, USA}
\author{Charles L. Bennett}
\affiliation{The Johns Hopkins University, Baltimore, Maryland, USA}
\author{Thomas M. Essinger-Hileman}
\affiliation{NASA Goddard Space Flight Center, Greenbelt, Maryland, USA}
\author{Haiquan Guo}
\affiliation{Universities Space Research Association, Columbia, Maryland, USA}
\affiliation{NASA Glenn Research Center, Cleveland, Ohio, USA}
\author{Tobias Marriage}
\affiliation{The Johns Hopkins University, Baltimore, Maryland, USA}
\author{Manuel A. Quijada}
\affiliation{NASA Goddard Space Flight Center, Greenbelt, Maryland, USA}
\author{Ariel E. Tokarz}
\affiliation{NASA Glenn Research Center, Cleveland, Ohio, USA}
\author{Stephanie L. Vivod}
\affiliation{NASA Glenn Research Center, Cleveland, Ohio, USA}
\author{Edward J. Wollack}
\affiliation{NASA Goddard Space Flight Center, Greenbelt, Maryland, USA}


\date{\today}

\begin{abstract}
Infrared-blocking, aerogel-based scattering filters have a broad range of potential applications in astrophysics and planetary science instruments in the far-infrared, sub-millimeter, and microwave regimes. This paper demonstrates the ability of conductively-loaded, polyimide aerogel filters to meet the mechanical and science instrument requirements for several experiments, including the Cosmology Large Angular Scale Surveyor (CLASS), the Experiment for Cryogenic Large-Aperture Intensity Mapping (EXCLAIM), and the Sub-millimeter Solar Observation Lunar Volatiles Experiment (SSOLVE). Thermal multi-physics simulations of the filters predict their performance when integrated into a cryogenic receiver. Prototype filters have survived cryogenic cycling to 4\,K with no degradation in mechanical properties. Measurement of total hemispherical reflectance and transmittance, as well as cryogenic tests of the aerogel filters in a full receiver context, allow estimates of the integrated infrared emissivity of the filters. Knowledge of the emissivity will help instrument designers incorporate the filters into future experiments in planetary science, astrophysics, and cosmology.
    
\end{abstract}


\maketitle 


\section{Introduction}
\label{sec:intro}  
Receivers common in the far-infrared, sub-millimeter, and microwave regimes require strong out-of-band infrared rejection to maintain cryogenic performance and desired signal-to-noise ratios. A range of strategies have been employed to realize thermal infrared (IR) blocking filters, which use reflective metal meshes, absorptive materials, and/or scattering particles using a variety of materials.\cite{ULRICH196737, ULRICH196765, ade_filters, Bock:95, Halpern, Munson:17, Timusk:81, choi, Whitbourn:85, Inoue_alumina_filter_2014, Takaku_alumina_filter_2021}

\begin{figure}[h!]
    \centering
    \includegraphics[width = \columnwidth]{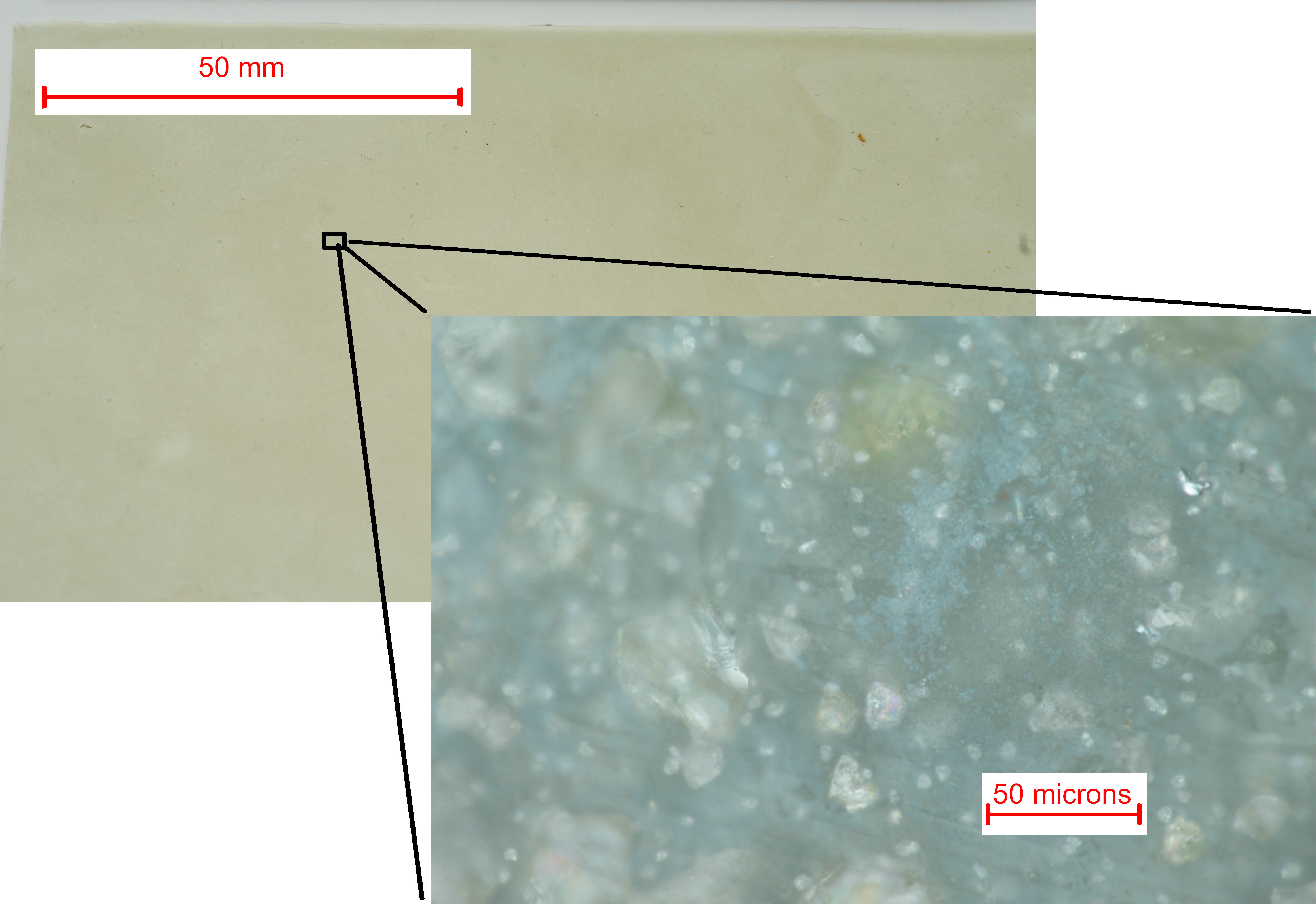}
    \caption{A composite image of a prototype diamond-loaded polyimide aerogel filter. The sample is a prototype formulation designed for a sub-millimeter or microwave instrument, approximately 0.2\,mm thick. Lower right image shows an optical microsope image of the aerogel. The pores of the aerogel are too small to be seen at this scale, but several different sizes of diamond particles are visible. For more details of the composition of the sample, see sample G in Table. \ref{table:params}}.
    \label{fig:FormulaG}
\end{figure}

Diamond-loaded polyimide aerogels are a good choice for infrared-blocking filters because of their excellent combination of physical and electromagnetic characteristics.~\cite{meador1, meador2, guo, guo2} A detailed discussion of the fabrication and characterization of diamond-loaded polyimide aerogel formulations for this application is available in Ref.~\citenum{Barlis:24}. An image of a prototype aerogel filter is in Figure \ref{fig:FormulaG}. These filters offer several benefits when compared to existing state-of-the-art IR filter technologies, including a tunable cutoff over a wide range of frequencies; high, broadband, in-band transmission; and strong rejection of out-of-band power across a broad range of frequencies. 

This paper demonstrates the unique characteristics of diamond-loaded polyimide aerogel filters compared to existing materials, shows their ability to meet specific instrument requirements, demonstrates their operation in an integrated cryogenic receiver, and constrains their emissivity. Specific instrument applications considered are for the Cosmology Large Angular Scale Surveyor (CLASS), the EXperiment for Cryogenic Large-Aperture Intensity Mapping (EXCLAIM), and the Sub-millimeter Solar Observation Lunar Volatiles Experiment (SSOLVE). These experiments span the microwave (CLASS),\cite{CLASS, CLASS2025, CLASS_Eimer_2024, CLASS_TEH_2014} sub-millimeter, (EXCLAIM)\cite{EXCLAIM_Overview_Switzer_2021, EXCLAIM, EXCLAIM_Forecast_Pullen_2023, EXCLAIM_Optics_TEH_2025} and far-IR (SSOLVE)\cite{SSOLVE} with applications to cosmology, galaxy formation, and planetary science, respectively. Developing filters for these mission applications, which place very different constraints on filter design and fabrication, allows an exploration of the versatility of aerogel scattering filters, as well as their limitations.

This paper is organized as follows: Section~\ref{sec:filter_comparison} summarizes the current state of the art for IR-blocking filters and places aerogel scattering filters in context; Section~\ref{sec:characterization} details optical characterization of materials and filters, including estimates of filter emissivity through measurement of total hemispherical transmission and reflection; Section~\ref{sec:fabrication} details the fabrication of large films (up to 50~cm wide) aimed at CMB receivers; Section~\ref{sec:mission_applications} provides an overview of filter design and describes the design and characterization of filters for the specific needs of the CLASS, EXCLAIM, and SSOLVE instruments; Section~\ref{section:thermal} describes modeling of the thermal performance of the filters in a CLASS receiver using \textit{COMSOL Multiphysics}; Section~\ref{sec:receiver_testing} details tests of large format aerogel filters in a CLASS receiver; and, we conclude in Section~\ref{sec:conclusion}.

\section{Comparison of Infrared-Blocking Filter Technologies}
\label{sec:filter_comparison}
\begin{figure}[h!]
    \centering
    \includegraphics[width = \columnwidth]{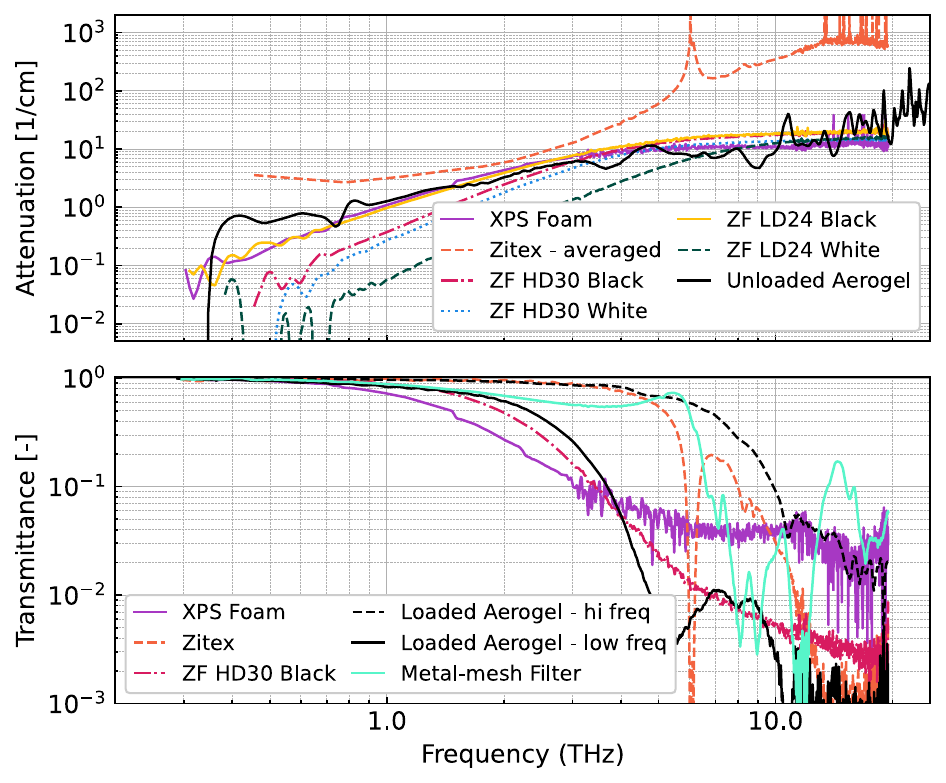}
   \caption{\textit{Top:} Comparison of the attenuation coefficient of a variety of infrared blocking filter materials. The Zotefoam (ZF) samples, XPS foam, and unloaded aerogels all have comparable attenuation. The data for the Zitex sample below 4\,THz are binned because the thinness of the sample produces significant etalon fringing in the measurement data. \textit{Bottom:} Comparison of the transmittance of different infrared filters. The Zotefoam filter and XPS foams are shown for a filter 3\,mm thick. The Zitex is shown for a filter that is 0.1\,mm thick. The metal-mesh filter is 0.004\,\micron{} thick. The low frequency aerogel is Formula G and the high frequency aerogel is Formula C from Table \ref{table:params}.}
    \label{fig:attenuation}
\end{figure}

We have developed a suite of diamond-loaded polyimide aerogel filters that offer more versatile and customizable optical performance in comparison to other materials and technologies used for IR rejection. Figure \ref{fig:attenuation} shows a comparison of the attenuation coefficient and the transmission properties for a variety of common infrared-blocking materials.

We compare the performance of the diamond-loaded polyimide aerogel filters to the following materials/technologies: 
\begin{itemize}
  \item Absorptive bulk materials
  \item Expanded polymer foams
  \item Porous polytetrafluoroethylene (PTFE)
  \item Inductive/capacitive metal-mesh filters.
\end{itemize}
Generally speaking, the electromagnetic properties of foams and bulk materials can be modeled using the Maxwell-Garnett approximation\cite{MaxwellGarnett1904} or other similar methods\cite{Sihvola2008, Looyenga1965}. The Maxwell-Garnett approximation treats a foam or dielectric mixture as a matrix with known dielectric function $\epsilon_{m}$ with inclusions with dielectric function $\epsilon_{i}$, with volume filling fraction $v_{ff}$. 

\begin{equation}
\epsilon_{eff} = \epsilon_{m} \frac{2v_{ff}(\epsilon_{i} - \epsilon_{m}) + \epsilon_{i} + 2\epsilon_{m}}{2\epsilon_{m} + \epsilon_{i} - v_{ff}(\epsilon_{i} - \epsilon_{m})}.
    \label{eq:maxwell-garnett}
\end{equation}
For foams and aerogels, the matrix can be treated as vacuum with small inclusions of bulk polymer(s). For the porous PTFE, the relationship is reversed and the material is modeled as bulk PTFE with inclusions of vacuum. More in-depth modeling of the electromagnetic properties of the diamond-loaded aerogels can be found in Ref. \citenum{Barlis:24}. Table \ref{table:foams} shows a summary of the physical and electromagnetic properties of the foams and similar polymer materials analyzed in this paper.
\begin{table}[h!]
\centering
\resizebox{\linewidth}{!}
{%
\begin{tabular}{|r|c|c|c|c|} 

 \hline
\textbf{Sample Source} & \textbf{Typical} & \textbf{Volume} & \textbf{Bulk Relative} & \textbf{Maxwell-Garnett} \\
\textbf{Polymer} & \textbf{Pore Size} & \textbf{Fraction} & \textbf{Permittivity} & \textbf{Permittivity}  \\

{} & $d_{pore}$ [\textmu m] & $f_{v}^{pore}$ [-] & $\epsilon^{bulk}_{r}$ [-] &  $\epsilon^{MG}_{r}$ [-]  \\
\hline
Plastazote LD24 (ZF LD24) &\multirow{2}{*}{500}  & \multirow{2}{*}{0.025} & \multirow{2}{*}{2.31 + 0.0004$i$} & \multirow{2}{*}{1.022 + 0.0000049$i$} \\
polyethylene (white \& black) & {}&{} & {} & {} \\
\hline
Plastazote HD30 (ZF HD30) &\multirow{2}{*}{400}  & \multirow{2}{*}{0.032} & \multirow{2}{*}{2.31 + 0.0004$i$} & \multirow{2}{*}{{1.029 + 0.000006$i$}} \\
polyethlyene (white \& black) & {}&{} & {} &{}\\
\hline
Zitex G-115&\multirow{2}{*}{1-2}  & \multirow{2}{*}{0.4} & \multirow{2}{*}{2.1 + 0.0004$i$} & \multirow{2}{*}{1.609 + 0.00021$i$} \\
PTFE & {}&{} & {} &{} \\
\hline
RL Adams Foam (XPS Foam) &\multirow{2}{*}{300}  & \multirow{2}{*}{0.03} & \multirow{2}{*}{2.53 + 0.034$i$} & \multirow{2}{*}{1.031 + 0.00046$i$} \\
polystyrene& {}&{} & {} &{} \\
\hline
\end{tabular}
}
\caption{A summary of the physical properties of the different foams and filter media measured. Bulk relative permittivities for polyethylene and PTFE are found in Ref. \citenum{goldsmith2002quasi} and polystyrene in Ref. \citenum{BIRCH1991}. Pore sizes are from manufacturer data sheets\cite{LD24, HD30}. Volume filling fractions for LD24 and HD30 are calculated using the manufacturer's given density and the bulk properties of polyethylene\cite{LD24, HD30}. Volume filling fraction for Zitex G-115 is specified by the manufacturer\cite{zitex-data}. Volume filling fraction for RL Adams foam is calculated using density values measured by Ref \citenum{petroff-thesis}. }
\label{table:foams}
\end{table}
The dielectric functions of different media can be measured in a few different ways. A common method is to fill pieces of rectangular waveguide with the sample media and measure it using a vector network analyzer (VNA). More details about this type of measurement are available in Ref. \citenum{Barlis:24}. Measurements can also be made in free space like in Ref. \citenum{Eimer2011}.

Some materials with relatively high transmission at long wavelengths but strong IR absorption have been used as filters. These materials include alumina, fluorogold, nylon, and polytetrafluoroethylene (PTFE).\cite{Halpern, Inoue_alumina_filter_2014, Takaku_alumina_filter_2021}
Filters using absorbing materials typically require anti-reflection (AR) coatings to reduce reflection from their surfaces. Such AR coatings add to the complexity of filter fabrication and can pose delamination risk in a cryogenic setting. Moreover, AR coatings are tuned to the instrument passband, limiting overall filter bandwidth. Multi-layer AR coatings increase bandwidth at the expense of additional fabrication complexity.

\begin{figure}[ht]
    \centering
    \includegraphics[width = 0.9\columnwidth]{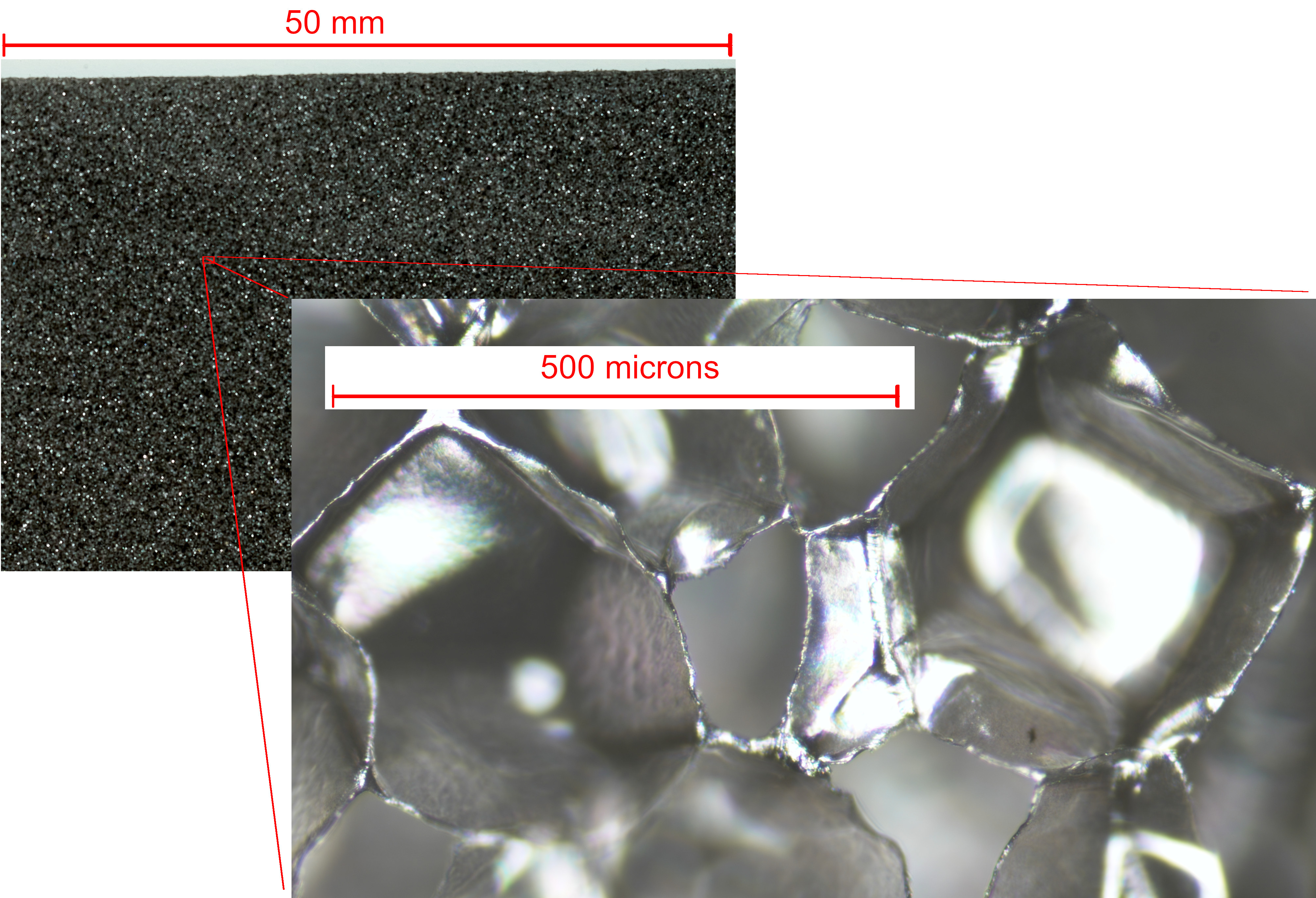}
    \includegraphics[width = 0.9\columnwidth]{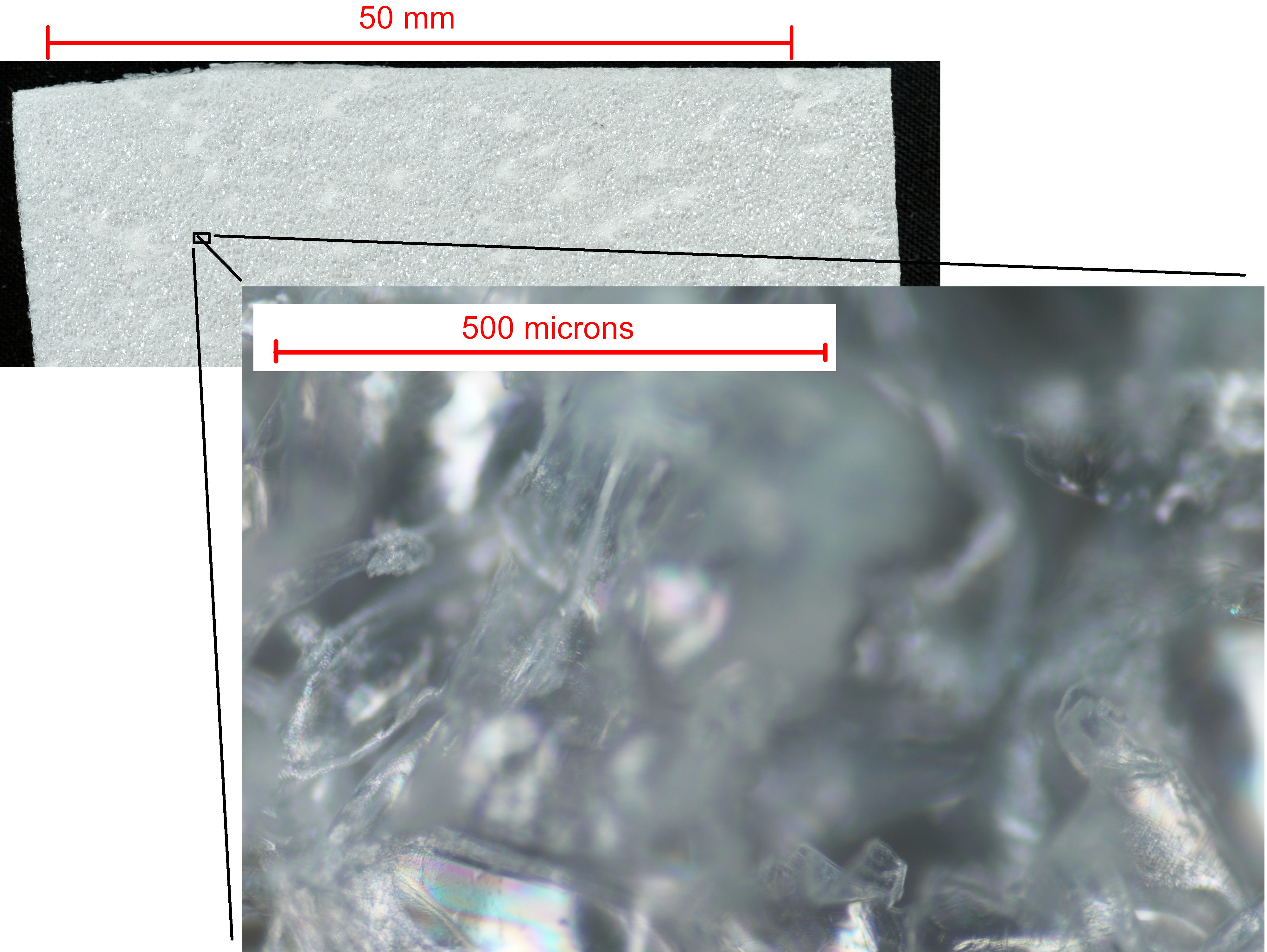}
    \includegraphics[width = 0.9\columnwidth]{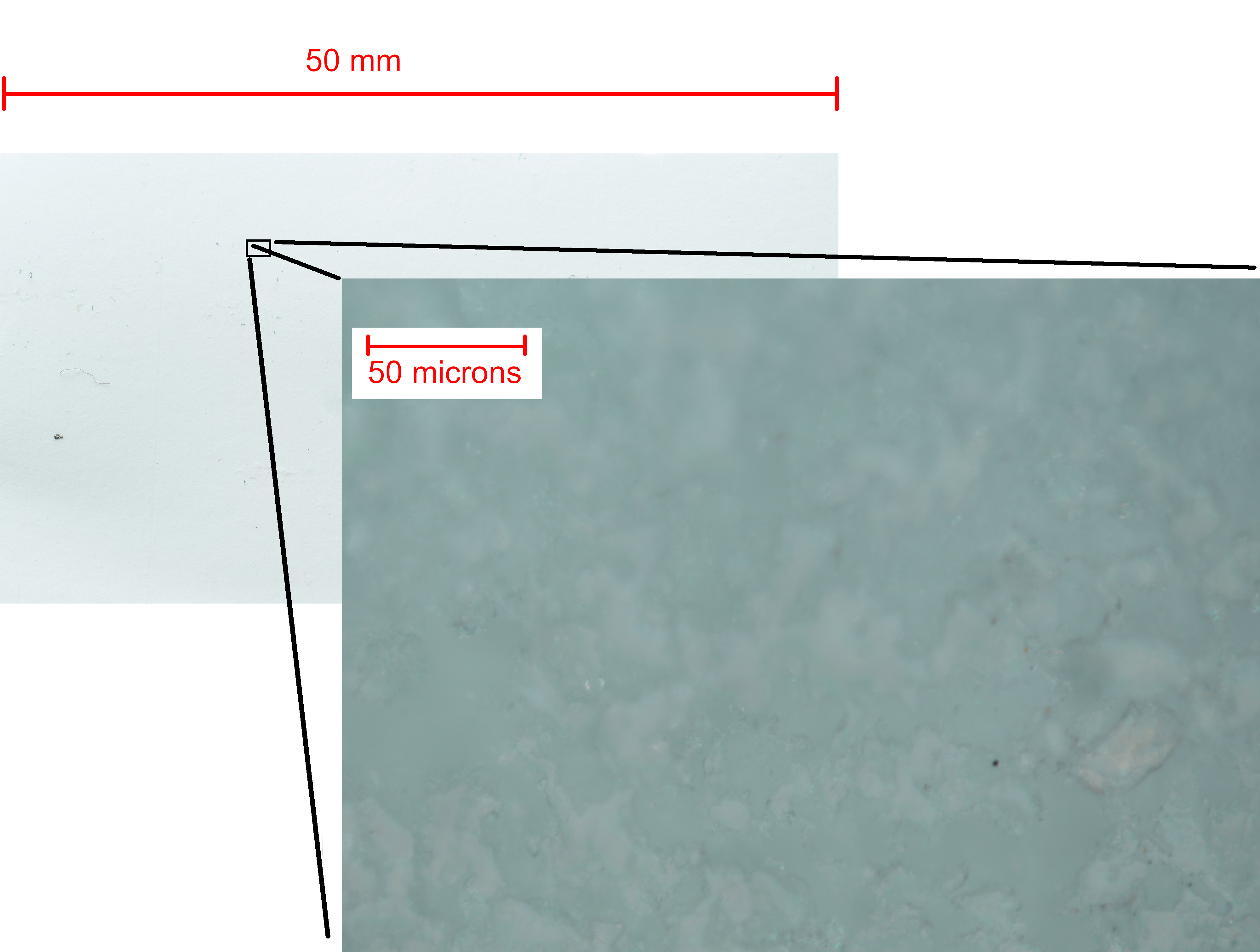}
    \caption{Three composite images of some of the commercial polymer-based products tested and compared to the diamond-loaded polyimide aerogel filters. Top composite is the black colored Zotefoam HD30. Middle composite image is XPS Model Airplane Foam. Bottom composite is Zitex G115. Additional images are available in Appendix \ref{image_appendix}. Images show both a macroscopic image of a roughly 50\,mm wide sample as well as a microscope images to show the detail of the substructure of the media. The foams have obvious cells and filaments but the 1-2\textmu m pores of the Zitex are too small to be seen with high magnification.}
    \label{fig:foams}
\end{figure}

To avoid the need for AR coatings, high-porosity materials with correspondingly low index of refraction are also widely used. Expanded polymer foams are a popular choice for infrared rejection. These include closed-cell polyethylene foams like Zotefoam Plastazote HD30 \cite{HD30} and LD24\cite{LD24}. The foams are readily commercially available in a variety of thicknesses, ranging from 3 to 100\,mm. Images of Zotefoam samples are shown in Figure \ref{fig:foams}. The HD and LD varieties have different densities and thus exhibit a range of cutoff frequencies. The foams are also available in white or black (and a variety of other colors) and the visible color corresponds to changes in infrared transmission properties. A comparison of the optical properties of expanded polymer foams and the diamond-loaded polyimide aerogel filters is shown in Figure \ref{fig:attenuation}. One substantial drawback to the expanded polyethylene foams is that a limited number of commercially-manufactured formulations are available. The pore cell size of a foam largely determines its spectral performance and cutoff frequency. Typical pore sizes for expanded polymer foams are typically $> 50$\,\micron, which strongly scatter wavelengths $\lambda \lesssim 100$\,\micron. This is desirable for some applications, but not others. The only consumer-adjustable property is therefore the thickness of the foam material, but this comes with additional considerations for cryogenic performance and mechanical stability. Additionally, many foams come with a skin or top layer of bulk plastic that must be removed to reduce reflection and transmission fringes from the bulk plastic.

The CLASS project employs a commercial extruded polystyrene (XPS) foam manufactured by R. L. Adams (formerly available from Model Airplane Foam),\cite{petroff-thesis} with foam pores typically between 200\,\micron{} and 300\,\micron. Figure \ref{fig:attenuation} shows the infrared transmission spectrum of the XPS foam. The foam comes from the manufacturer with a face sheet, which is a thin layer of bulk polymer on one side that must be removed before use on CLASS. The total thickness of a piece is approximately 6\,mm. Removing the face sheet without damaging or tearing the foam material itself can be difficult and time consuming. Additionally, because it is a commercial off-the-shelf (COTS) foam, its spectral performance cannot be tuned or adjusted, except that thicker foams can be milled or cut down to smaller thicknesses to increase overall transmission. The CLASS collaboration milled the foam pieces down to approximately 1\,mm thick sheets for use in the cryogenic receivers. The foam has a much more gradual cutoff and lower frequency $-3$\,dB point than others, which is acceptable because the CLASS observations only need to extend up to 220\,GHz. 

Microporous PTFE, such as Zitex\cite{zitex-data} and Porex,\footnote{Porex Filtration Group, https://www.porex.com/products/} is also a promising candidate material for use as an IR-blocking filter. A range of formulations are commercially available with typical pore sizes down to 1\,\micron. Scattering from the small micropore structure produces filter cutoffs at higher frequencies/shorter wavelengths than for expanded foams; however, the intrinsic absorption of PTFE at THz frequencies limits use of these materials to the microwave and sub-millimeter. Similar to expanded polymer foams, the properties of microporous PTFE are not particularly tunable to the desired wavelength range of the end user. It can be procured in a variety of thicknesses, but the cutoff frequency is largely set by the pore size, high index of refraction (n $\simeq 1.44$), and high loss of PTFE in the THz. 

Another common filter technology employs patterned metal mesh on a polymer substrate, currently available from QMC Instruments Ltd. and Cardiff University.\cite{ade_filters} These filters are thin polypropylene membranes with inductive and capacitive metalized films lithographed onto the surface. An in-depth discussion of the physics behind metal mesh filters can be found in Ref.\citenum{metalmeshfilters}. The feature size and feature spacing set the high-pass, low-pass, or band-pass frequencies. Multiple layers can be bonded together, typically through heat-pressing with polyethylene layers, to produce more complex filter arrangements. These filters act somewhat differently than foams and polymers because the conductive patterning, instead of the physical properties of the polymer like the index of refraction or bubble size, is what determines the electromagnetic behavior of the filter. The carrier membrane film only needs to be thick enough to support the manufacturing and mechanical installation of the filters. 

\begin{figure}[h!]
    \centering
    \includegraphics[width = \columnwidth]{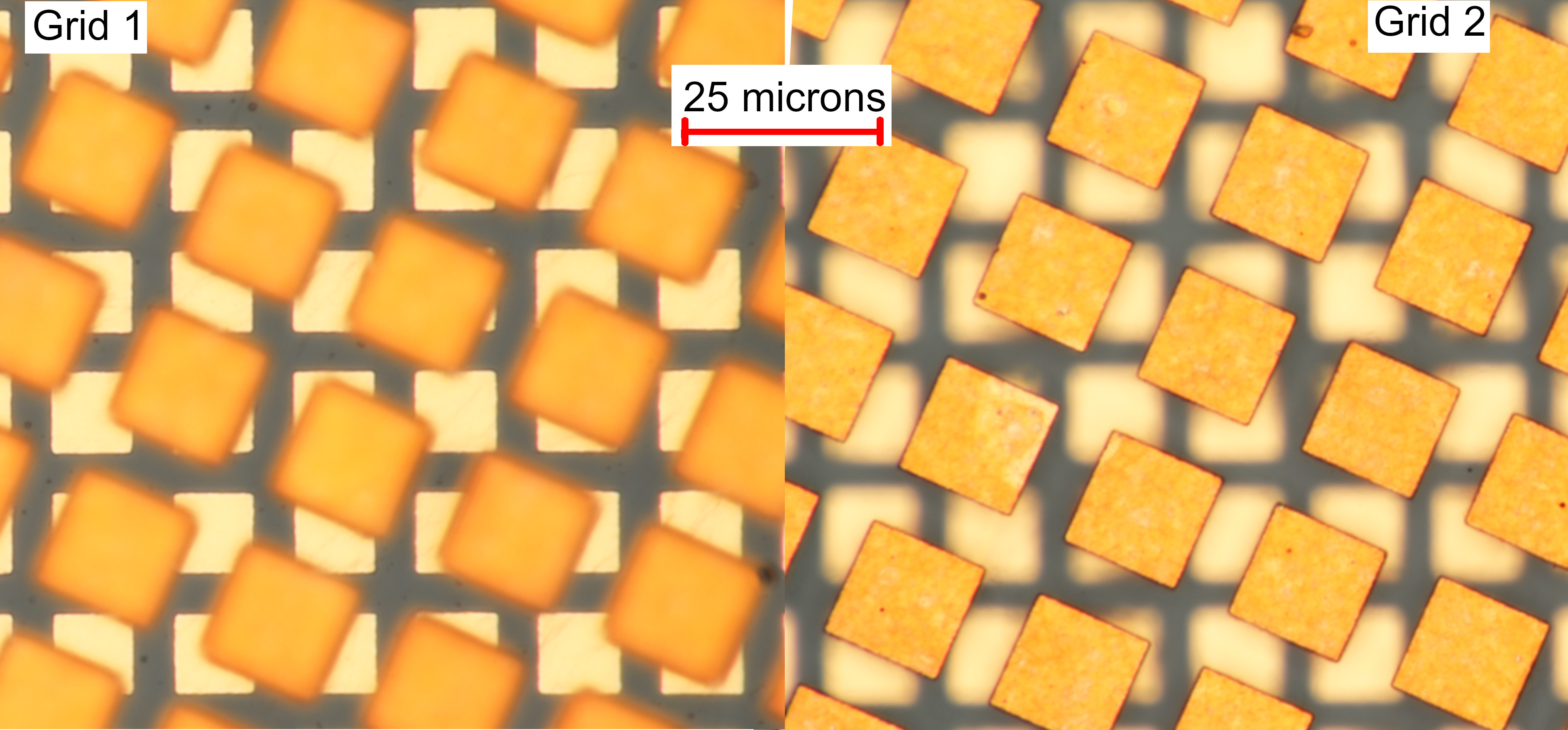}
    \caption{Optical microscope images of the metalized metal mesh grids on a Cardiff University/QMC filter \cite{IR_shaders}. Grid 1 are the smaller yellow squares with gray borders between them. Grid 2 are larger orange squares that are tilted relative to the smaller yellow squares. The two metal meshes are deposited on a thin, 3.3\,\micron{} polypropylene film. Grid 2 has a square size of approximately 17\,\textmu m with a center-to-center spacing of 25\,\textmu m. Grid 1 has a square size of approximately 11\,\micron{} with a center-to-center spacing of approximately 17\,\micron. The filter's optical transmittance is shown in Fig. \ref{fig:attenuation}. }
    \label{fig:cardiff-composite}
\end{figure}

One specific metal-mesh filter architecture is designed to reject infrared light near the peak emission of a 300\,K (or other specified temperature) blackbody. These filters are called ``infrared shaders'' \cite{IR_shaders} and are constructed of a single thin, $\sim3$\,\textmu m polypropylene film with capacitive grids on both sides. These grids can be seen in microscope images in Figure \ref{fig:cardiff-composite}. They have feature sizes that are approximately 20\,\micron{} wide with spacing that is less than 10\,\micron.

Scattering filters like foams and aerogels reject unwanted light somewhat differently than metal-mesh filters. Metal-mesh filters are mostly specular and reflect light backwards through an optical system. This reflected light must be able to pass backwards through the preceding optical elements. Foams and scattering filters will reflect light more isotropically into the surrounding environment. These considerations must be taken into consideration when desiging an instrument around a certain filter type.

\section{Optical Characterization of Materials and Filters}
\label{sec:characterization}
The optical response of the different filter substrates was measured with a Bruker 125HR Fourier Transform Spectrometer (FTS). For more information about the FTS measurement setup, see Barlis, et al, Ref. \citenum{Barlis:24}. A comparison of the transmission of a Cardiff-manufactured metal-mesh filter, Zitex G115, one Zotefoam sample, the CLASS XPS foam, and an aerogel filter of a similar frequency range is seen in Figure \ref{fig:attenuation}. The metal-mesh transmission curve shows significantly more structure and in-band non-uniformity than the aerogel filter over a similar wavelength range as well as more ``blue light leak'' or transmission beyond the band edge near 10\,THz.

Our diamond-loaded polyimide aerogel filters offer several benefits when compared to other available solutions. Firstly, their cutoff frequencies are tunable by varying the size and loading density of the embedded diamond particles. Secondly, they have a low index of refraction ($n \approx1.15$), which eliminates the need for AR coatings. Thirdly, they have high in-band transmission while maintaining low out-of-band transmission. Finally, they are manufacturable in large sizes (e.g. >\,50\,cm) to enable use in instruments with high throughput.

\subsection{Total Hemispherical Measurements of Diamond-loaded Polyimide Aerogel Filters and Other Media}

\begin{figure}
    \centering
    \includegraphics[width=\columnwidth]{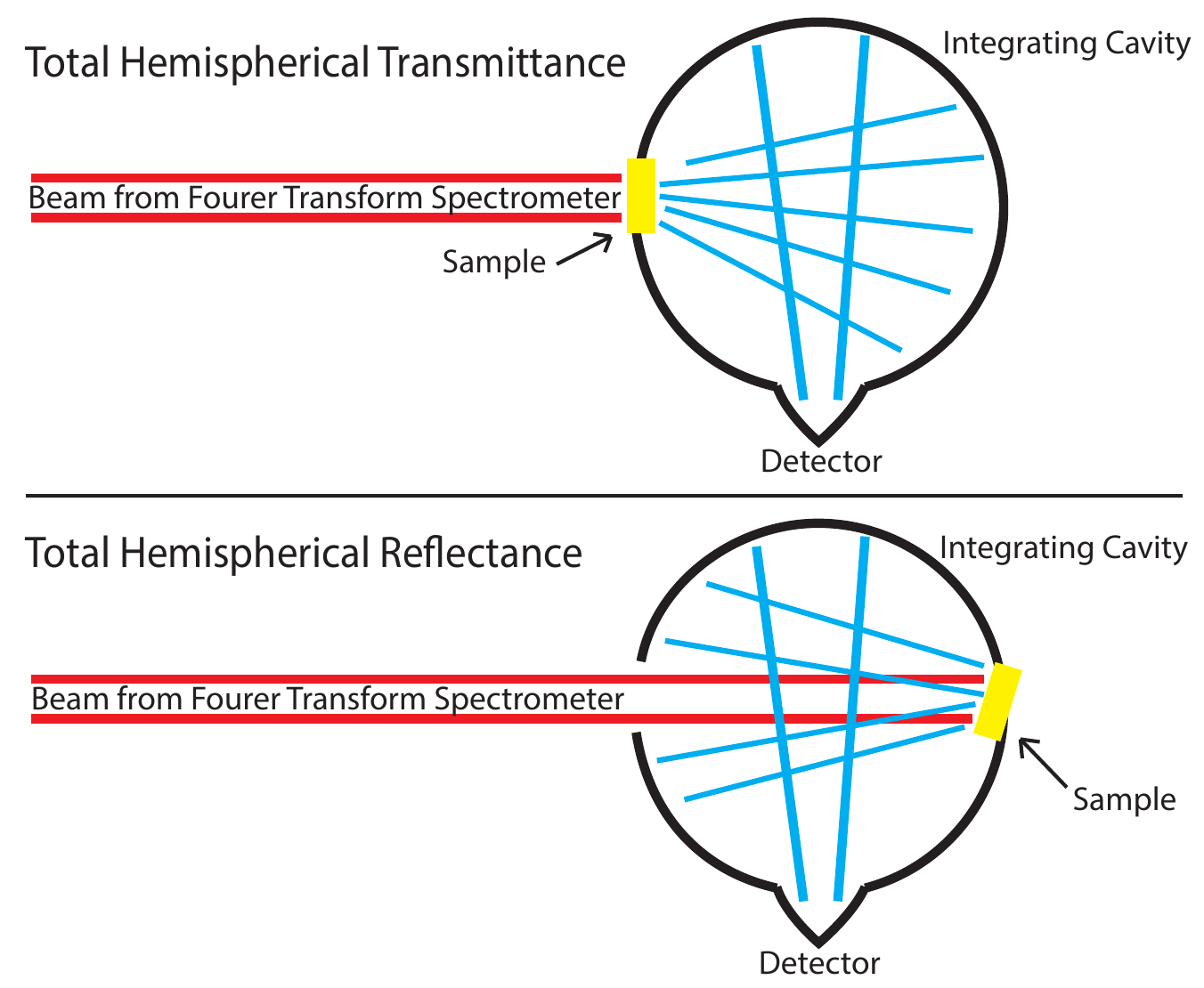}
    \caption{A cartoon schematic of an integrating sphere assembly designed to measure both total hemispherical transmittance and reflectance. The upper configuration shows the setup to measure transmittance while the lower configuration shows the setup to measure reflectance. To measure transmittance, the beam from the FTS is sent through the sample into the integrating cavity. The light from all angles bounces around inside the sphere and is captured by the detector. To measure reflectance, the beam from the FTS is sent in to the integrating cavity and reflected off the sample at non-normal incidence. Light reflected from the sample at all angles is captured by the integrating sphere and measured by the detector. }
    \label{fig:integrating-sphere}
\end{figure}

Measurements of the total hemispherical reflectance and transmittance of a filter allow the characterization of how much light is reflected by or transmitted through a sample over all angles, instead of just the light along the optical axis of the incoming beam. Figure \ref{fig:integrating-sphere} shows a drawing of the integrating sphere setup used to measure the total hemispherical transmittance and reflectance. 

\begin{figure}[h!]
    \centering
    \includegraphics[width = \columnwidth]{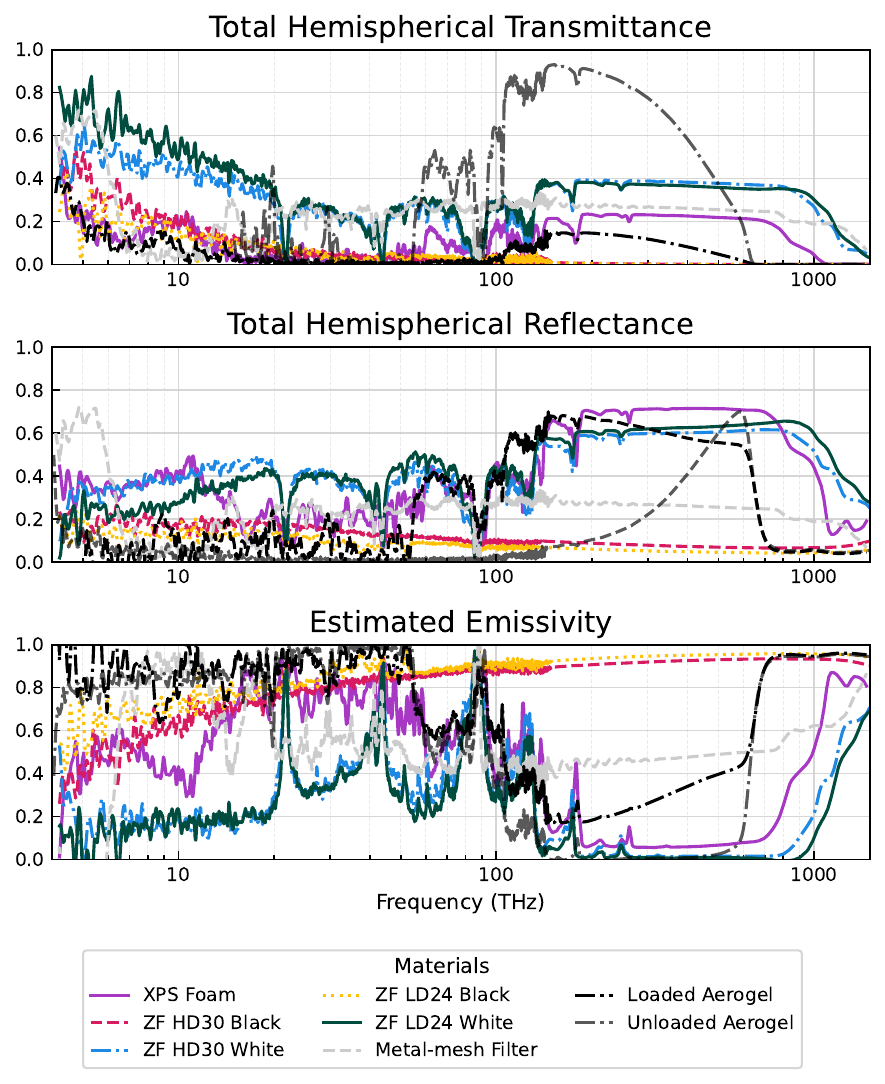}
   \caption{Comparison of the measured total hemispherical transmittance (top panel), measured total hemispherical reflectance (middle panel), and estimated emissivity (bottom panel) of different infrared filters. The Zotefoam filter and XPS foams are shown for a filter 3\,mm thick. The Zitex is shown for a filter that is 0.1\,mm thick. The diamond-loaded polyimide aerogel filter is an EXCLAIM candidate formulation, Formula G from Table \ref{table:params}, and is 1.3\,mm thick. The unloaded aerogel is 1.4\,mm thick. The metal-mesh filter is an infrared shader 3.3\,\textmu m thick. The white LD24 and HD30 samples show higher transmission than the other samples, which leads to a lower calculated emissivity.}
    \label{fig:total_hemi}
\end{figure}

Diamond-loaded polyimide aerogel filter samples along with other commercially prepared foam samples were measured using a Bruker FTS equipped with an integrating sphere accessory. Samples were mounted to the integrating sphere and illuminated with a beam of light. The scattered reflected light was collected by the integrating sphere and measured by the detector. An analogous measurement was made to find the total hemispherical transmittance. Because these measurements were made with an FTS, transmittance and reflectance data are available as a function of frequency/wavelength of light. Figure \ref{fig:total_hemi} shows the total hemispherical data for the commercially available foam products as well as Recipe G of the EXCLAIM formulation of the diamond-loaded aerogel filters. 

An additional benefit of making the total hemispherical measurements is that the emissivity can be estimated by subtracting the sum of the reflectance and transmittance from unity:
\begin{equation}
    \epsilon \simeq 1  - T - R.
    \label{eqn: emissivity}
\end{equation}
Figure \ref{fig:total_hemi} shows the estimated emissivity for the various samples across a broad range of frequencies. The white Zotefoam samples have a substantially lower estimated emissivity than the black foams, the polystyrene foam, and the aerogel filter. The very structured emissivity data below 100\,THz is not a real frequency-dependent emissivity. These transmittance measurements are not expected to be the same as the measurements in Figure \ref{fig:attenuation} because the total hemispherical measurements are diffuse and the measurements in Figure \ref{fig:attenuation} are specular. 

\section{The Fabrication of large-scale Aerogel Filters}
\label{sec:fabrication}
The increasing throughput requirements of astrophysics experiments have driven increased demand for windows, filters, and other optical elements that are 30\,cm in diameter or larger. To meet this demand we have developed the ability to produce large diamond-loaded polyimide aerogel filters that are over 50\,cm wide. For additional details on the synthesis of the polyimide aerogel films, see Ref. \citenum{Barlis:24}.

The most widely practiced procedure in polyimide synthesis is the two-step polymerization method which involves an intermediate step forming a poly(amic acid) along the reaction pathway to forming the polyimide. Creating a polymer aerogel using this method requires the use of sol-gel polymerization by combining bi-functional monomers and solvent to form a colloidal solution forming a dilute polymer network within the solution, leading to gelation. The gel networks created by sol-gel method become the aerogel skeletal structure once the liquid portion is removed, creating the aerogel. 

To create a polyimide aerogel, equimolar amounts of dianhydrides and diamines, in a polar aprotic solvent (n-methyl-2-pyrrolidone (NMP)) first form a poly(amic) acid precursor, followed by chemical imidization utilizing a base catalyst (triethyl amine) and a water scavenger (acetic anhydride) and then cross-linked with a multifunctional monomer. 

The diamines, 2,2’-dimethylbenzidine (DMBZ) and 4,4'-Bis (4-aminophenoxy) biphenyl (BAPB), are first added to NMP to create a colloidal solution (sol) of suspended monomers in solvent. To this solution the dianhydride, biphenyl-3,3',4,4’-tetracarboxylic dianhydride (BPDA), is added in excess to react with the diamine to create the soluble amic acid oligomers with anhydride endcaps. The triamine, 3,5-triaminophenoxybenzene (TAB), is added to create a crosslinked polymer network followed by chemical imidization to initiate ring closure of the amic acid to form a crosslinked polyimide gel. The polyimide gel films undergo supercritical fluid extraction using liquid CO$_{2}$ to remove the solvent and obtain the final rigid aerogel films. 

To make a large sample film that is 460\,mm x 460\,mm square with 1\,mm thickness, 5.94\,g BAPB and 19.61\,g BPDA were mixed together and stirred into 326.03\,mL NMP. The whole mixture was stirred for 30 minutes. Then 10.27\,g DMBZ, mixed in 20\,ml NMP was added in the mixture until the sol cleared. Subsequently, 0.57\,g TAB mixed in 10\,mL NMP was added to the sol. At this point in the procedure, diamond particles with different sizes and loading amounts were added. After the diamond addition, 50.42\,mL of acetic anhydride and 4.65\,mL of triethylamine were added. Following this sol-gel synthesis, the diamond-loaded film underwent supercritical CO$_{2}$ fluid extraction to produce the aerogel.

We investigated three different methods for fabricating large area films. The first was a vertical casting mold, which is constructed of two rigid pieces of plastic separated by spacers of a prescribed thickness. After stirring for 15 minutes, the sol was poured into a vertical mold 52.5\,cm x 52.5\,cm with the spacer thickness 1.2\,mm, thus the opening is 52.5\,cm x 1.2\,mm. Gravity pulls the diamond particles along the gravity vector during curing, however, because of the long vertical dimension of the mold, the dispersion of large and small particles is not as substantial as it is with horizontal casting methods. The inhomogeneous sections of the top and bottom edges can be cut away and the middle homogeneous section can be used for the final product.

The second method to cast large films uses a large tabletop casting machine with a doctor blade that is held at a specific gap above the flat surface. Sol was evenly spread using the doctor blade set with a 1.55\,mm gap and cast on a polyethylene terephthalate (PET) carrier film of a tabletop tape casting machine (Warehouse Inc.) with a casting speed 0.03\,m/s. The resulted gel film was washed with 50\% NMP + 50\% acetone, then four times of acetone daily. The doctor blade produces more uniform thickness films than casting, however larger diameter diamond particles settle out to one side due to gravity. 

\begin{figure*}[ht!]
    \centering
    \includegraphics[width = 0.33\textwidth]{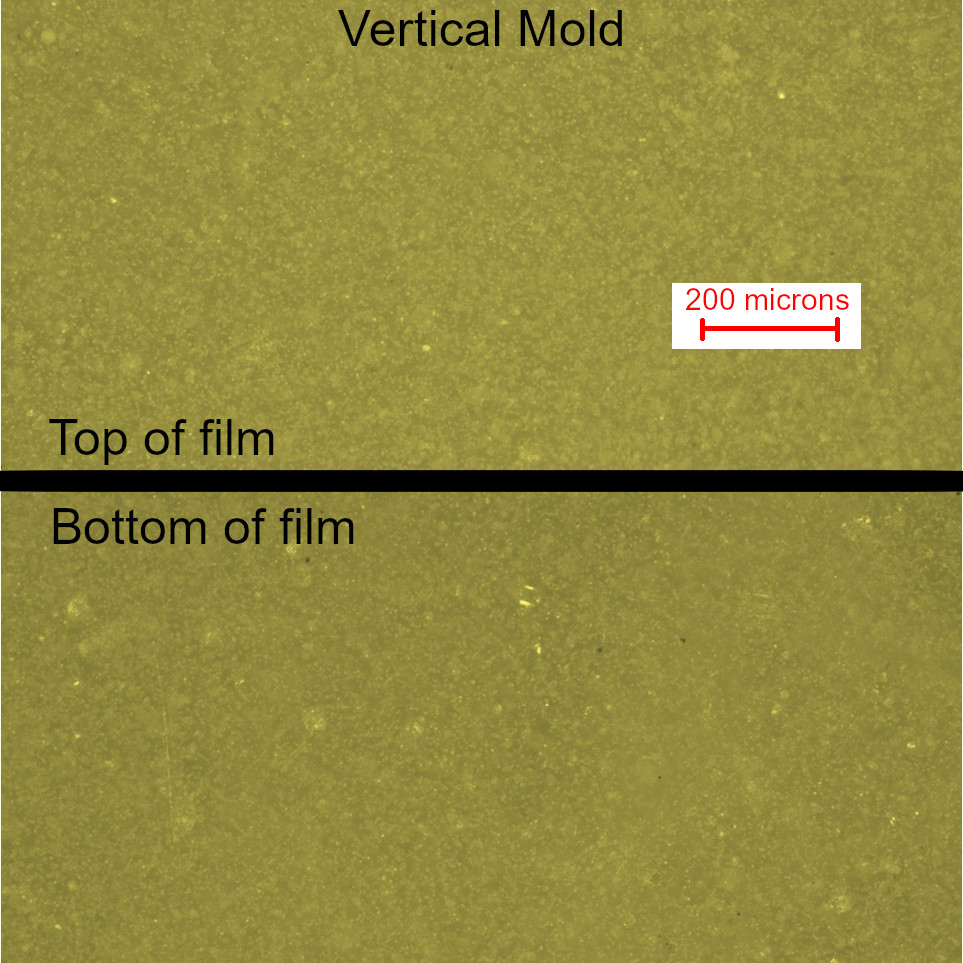}
    \includegraphics[width = 0.33\textwidth]{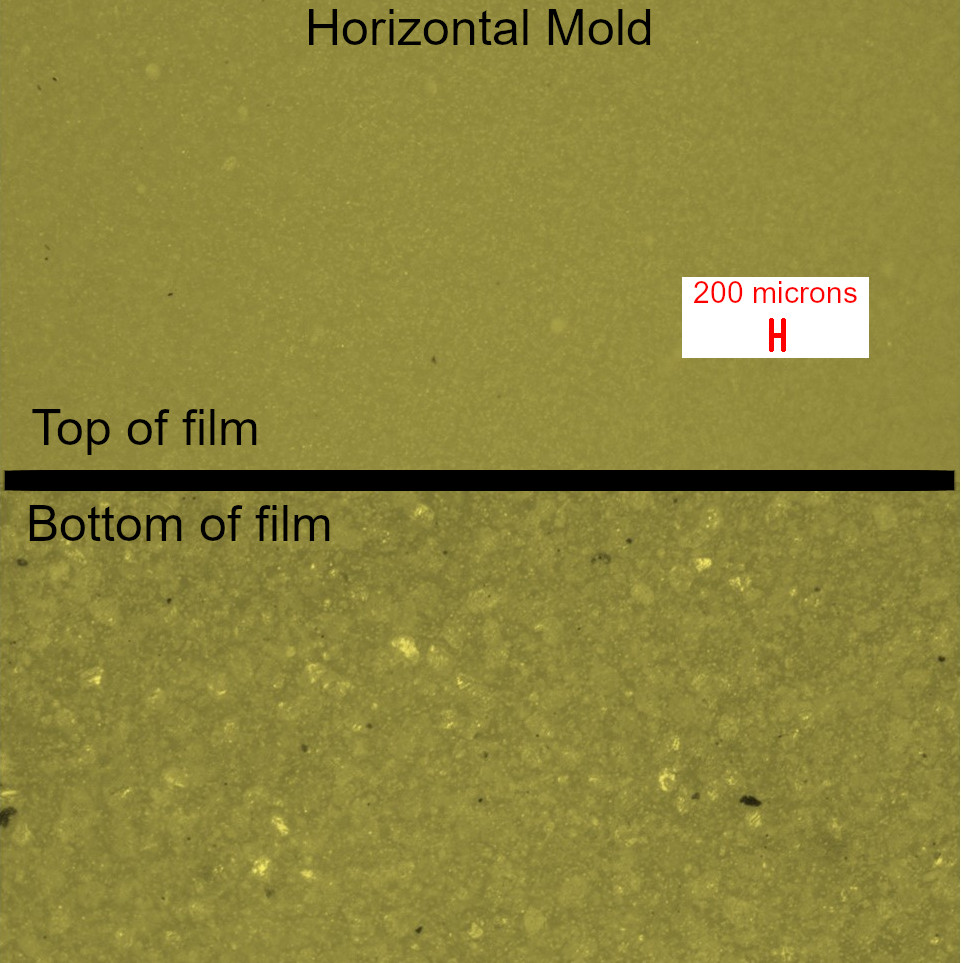} 
    \includegraphics[width = 0.33\textwidth]{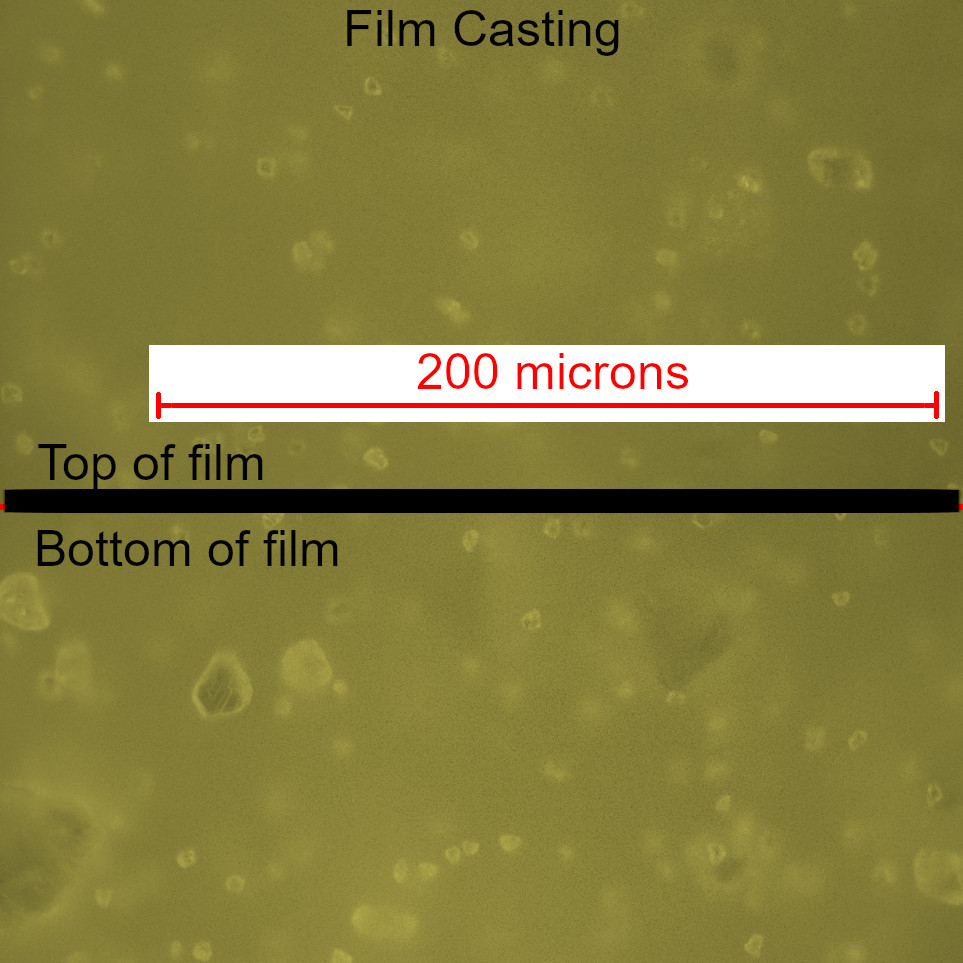} 
    
    \caption{Images comparing the top and bottom of different large diameter films made with the three different methods. The left image shows a sample from the vertical mold, the middle image shows a sample from the horizontal mold, and the right image shows a sample from the film casting table. Note the difference in the amount of settling of the larger diamond particles in the three different molding/casting methods. The vertical mold shows a similar distribution of small and large particles on the top and bottom of the film. The horizontal mold and the casting table show a larger number of larger diamond particles on the bottom of the films, due to the gravity vector pulling the particles downward during curing.}
    \label{fig:top-bottom-comparison}
\end{figure*}

A horizontal mold similar to the vertical mold was also used. Like the doctor blade method, the larger diameter particles settle out due to gravity. As seen in the optical microscope image of the filter samples in Figure \ref{fig:top-bottom-comparison}, the top side of the film using horizontal mold has fewer diamonds compared to the bottom of the film. Using the vertical mold or film casting methods, the diamond particles were dispersed in the films more homogeneously than in films made with the horizontal mold.

Film homogeneity can be judged by looking at the color of the film after it is dried. Areas with substantially fewer particles will be lighter yellow; areas with more particles will be darker yellow or even a greenish-yellow. The inhomogeneities are typically caused by the largest particles, especially with sizes greater than 100\,\micron. The particles 3-6\,\micron, 10-20\,\micron, and 40-60\,\micron{} have fewer problems, but the 250-350\,\micron cause the most inhomogeneities. Five different samples were taken from the center and the four corners of a large, $\sim 50$\,cm, film, with the corner samples being roughly 50\,mm away from the edge of the film. The samples were characterized on the FTS and found to be uniform within the error of the measurements. Removing a 50\,mm border from a 50\,cm film results in an approximately 60\% yield.

Our current method of drying is rolling the filters and then placing them in a supercritical dryer. The length of the film can be longer, but diameter will then be determined by the depth of the dryer chamber, the chamber at Glenn Research Center (GRC) can accommodate and approximately 55\,cm sample for reasonable drying. It can be scaled up to 1\,m or larger with a larger chamber.

The desired thickness of a film or filter is defined by the target application or frequency range. We made 0.2\,mm and 1\,mm for the above applications. Generally, the thinnest filter prepared by casting method at GRC is about 0.1\,mm to make it easy to peel off from the PET film. Very thick filters have other considerations. Depending on the casting method, thicker may result in more inhomogenous filters as the differently sized particles sink at different rates through the larger depth of the film. We also developed layer-by-layer casting to get laminated films, which could be used to get very thick films with a more homogeneous distribution of particles, or a specific controlled distribution of particle sizes through the thickness of the film.

For large diameter and large thickness filters, they are also more difficult to roll to be put inside the supercritical dryer. Depending on the formulation, some samples about 4-5\,mm thick can still flex about $180^{\circ}$ without breaking.

\section{Designing filters for specific mission requirements}
\label{sec:mission_applications}
Since our aerogel filter technology offers the opportunity to control the base aerogel chemistry, cutoff frequency, and in-band transmission, we are able to design filters to meet specific mission requirements. During the development and prototyping of filters, filters for three different instruments were designed, fabricated, and characterized. 

Aerogel filter samples were also tested for their cryogenic durability and survivability. Many experiments and instruments observing the THz through millimeter-wave regimes use cryogenic detectors requiring optics cooled to below 77\,K, and in some cases below 4\,K. Samples were repeatedly immersed in liquid nitrogen to test thermal shock performance. Filter samples were also installed inside of a cryogenic testbed and repeatedly cycled down to 4\,K from room temperature, as described in Section \ref{section:thermal}. While the filter mount must be designed to accommodate the CTE-mismatch between the polymer-based aerogels and other metallic or semi-metallic components, our polyimide aerogel filters are robust against thermal shock and cryogenic cycling. Because the primary component of the aerogel filters is a polymer plastic, polyimide, the filters are mounted in rings made of Delrin plastic. Delrin was chosen because it is known to work well at cryogenic temperatures. These plastic rings are screwed into aluminum filter mounts, so the CTE mismatch between the aluminum and the plastic is taken up by over-sized clearance holes on the plastic filter rings, not the aerogel filters themselves.

\subsection{Aerogel Filter Design Principles}
\begin{figure}[h!]
    \centering
    \includegraphics[width = \columnwidth]{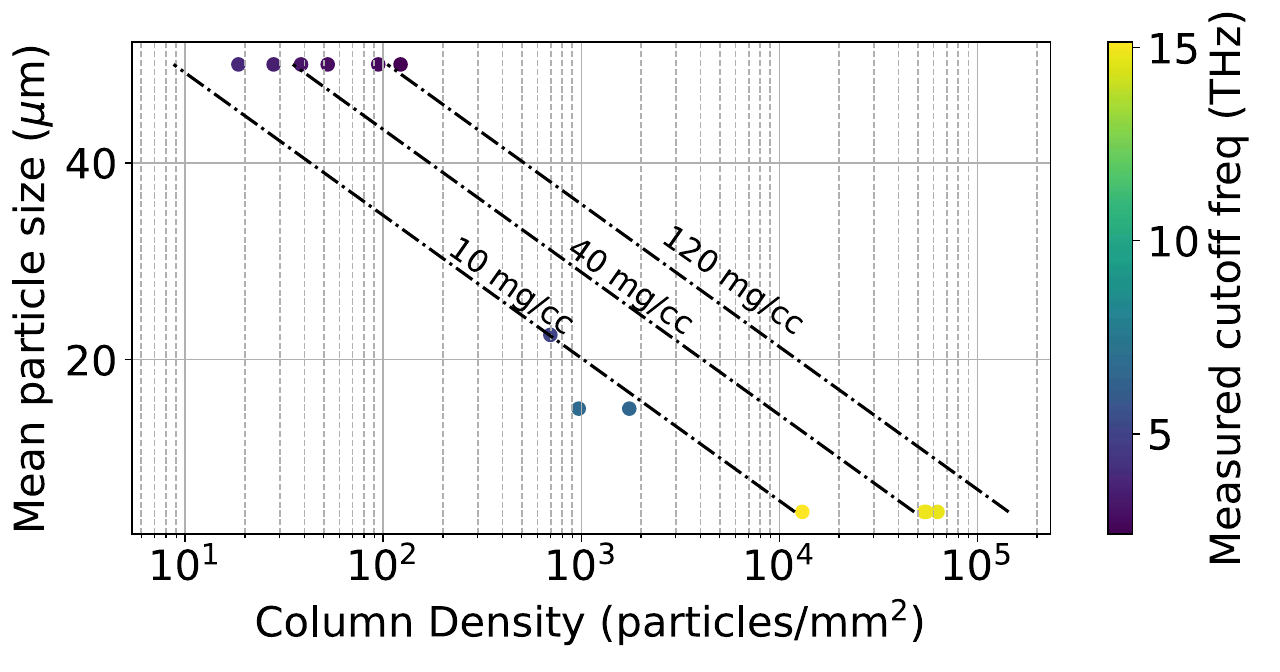}
    \caption{An overview of the diamond-loaded polyimide aerogel filter parameter space; diamond particle size, loading density (parameterized as column density), and subsequent measured cutoff frequency \cite{Barlis:24}. Particle size has the strongest effect on cutoff frequency but varying the loading density can allow for fine-tuning of the desired cutoff frequency. Diagonal dot-dashed lines show lines of constant column density for a giving loading density (10\,mg/cc, 40\,mg/cc, and 120\,mg/cc) and a 0.2\,mm thick sample.}
    \label{fig:aerogel_parameter_space}
\end{figure}

The diamond particle size and loading density are varied in order to tune the filter to the desired cutoff frequency. Particle size has the strongest effect on cutoff frequency. Larger particles have a lower cut-off frequency and smaller particles have a higher cut-off frequency. Diamond powders used typically have a Gaussian size distribution with the named range, e.g. 3-6\,\micron, representing the 3-$\sigma$ size range. Polyimide aerogel filters have been fabricated with a range of diamond loading densities ranging from 20\,mg/cc up to 100\,mg/cc. Loading densities significantly above 100\,mg/cc prevent proper polymerization and curing of the base aerogel. The base aerogels have a density of approximately 130\,mg/cc. Figure \ref{fig:aerogel_parameter_space} shows a summary of the parameter space generally available to us when designing and manufacturing diamond-loaded filters. For a given particle size distribution, a higher loading density creates a filter with a lower cut-off frequency and vice versa. In general, the scattering cross-section of the diamond particles goes as the square of the particle radius (area $\propto r^2$) and mass density goes as the cube of particle radius (volume $\propto r^3$). In order to capture the effect of particle radius and loading density, a column density can be calculated, that represents the total area of scatterers per unit area. Multiple particle size distributions and densities can be mixed together, e.g. primarily using a higher density of larger particles to set the primary cut-off frequency and mixing in lower densities of smaller particles to control high frequency leakage. 

\begin{table}[h!]
\centering
\resizebox{\linewidth}{!}
{%
\begin{tabular}{|c|c|c|} 

 \hline
 \textbf{Formula} & \textbf{Diamond} & \textbf{Target} \\ 
 \textbf{Name} & \textbf{loading}& \textbf{Experiment}\\
 \hline
A & 40-60\,\micron{} at 50\,mg/cc & SSOLVE \\
\hline

\multirow{2}{*}{B}  & 10-20\,\micron{} at 60\,mg/cc  & \multirow{2}{*}{SSOLVE} \\
{} & 3-6\,\micron{} at 20\,mg/cc &{} \\
\hline

\multirow{2}{*}{C} & 10-20\,\micron{} at 60\,mg/cc & \multirow{2}{*}{SSOLVE}\\
{} & 3-6\,\micron{} at 40\,mg/cc & {} \\
\hline

D & 3-6\,\micron{}  at 50\,mg/cc& SSOLVE \\
\hline

E & 3-6\,\micron{}  at 80\,mg/cc& SSOLVE \\
\hline

\multirow{2}{*}{F} & 40-60\,\micron{} at 40\,mg/cc & \multirow{2}{*}{EXCLAIM/CMB}\\
{} & 3-6\,\micron{} at 20\,mg/cc & {} \\
\hline

\multirow{3}{*}{G} & 40-60\,\micron{} at 40\,mg/cc & \multirow{3}{*}{EXCLAIM/CMB}\\
{} & 10-20\,\micron{} at 20\,mg/cc & {} \\
{} & 3-6\,\micron{} at 20\,mg/cc & {} \\
\hline

\multirow{2}{*}{H} & 40-60\,\micron{} at 60\,mg/cc & \multirow{2}{*}{EXCLAIM/CMB}\\
{} & 3-6\,\micron{} at 20\,mg/cc & {} \\
\hline

\end{tabular}
}
\caption{A summary of prototype diamond-loaded polyimide aerogel filters designed to meet specific mission needs.  A variety of filter loading schemes were fabricated and their transmission data were measured using a Bruker Fourier Transform Spectrometer (FTS). The SSOLVE mission requires good rejection of higher frequency THz light but greater than 90\% transmission at 2.5\,THz. A millimeter or sub-millimeter mission like a cosmic microwave background instrument or a line intensity instrument require high throughput below 1\,THz but strong rejection of higher frequency infrared light to aid in cryogenic performance of the instrument.}
\label{table:params}
\end{table}

\subsection{Terahertz Applications}
One application of loaded aerogel filters is in SSOLVE, a CubeSat mission concept to place a small satellite on the surface of the moon.~\cite{SSOLVE_Livengood_2020, SSOLVE} The SSOLVE instrument observes in three narrow bands centered around HDO, H$_2$O, and OH lines at frequencies of 23\,GHz, 533\,GHz, and 2.51\,THz, respectively. Using the Sun as a backlight, SSOLVE would search for absorption lines in the solar spectrum from these volatiles rising from the lunar regolith. This requires that out-of-band light from the Sun not cause excessive temperature fluctuations of the instrument, including a calibrator target. The aerogel filter for SSOLVE thus needed high ($>94$\%) transmission up to 2.51\,THz, while blocking $>95$\% of the integrated solar luminosity. This required a very thin ($\sim 200$\,\micron) filter over a 4\,cm aperture with small scattering particles.

\begin{figure}[h!]
    \centering
    \includegraphics[width = \columnwidth]{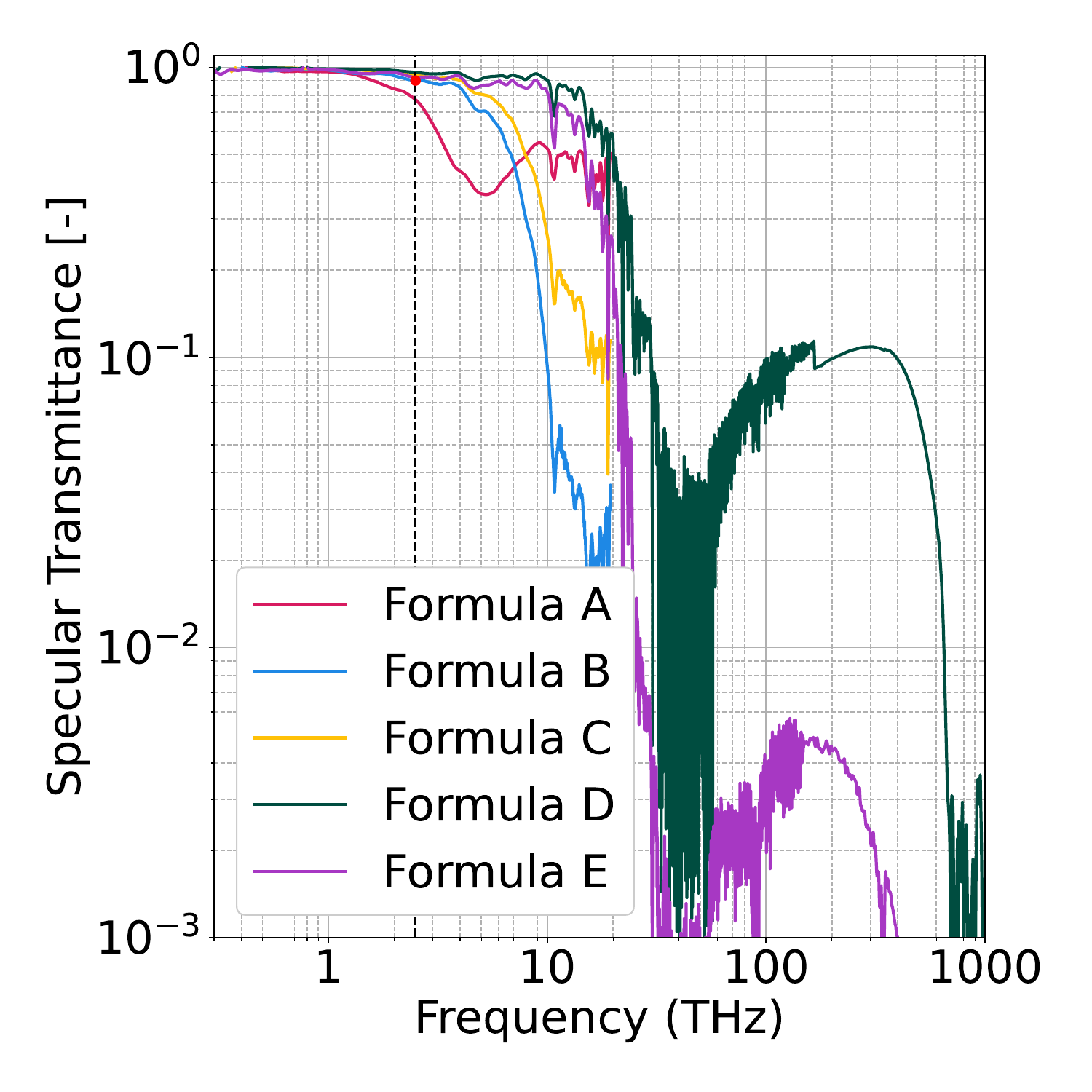}
   \caption{Transmission data for five different diamond-loaded polyimide aerogel filters for the SSOLVE mission. Filter loading schema are in Table \ref{table:params}. Note that measurement data for Formulas A, B, and C only extends to approximately 20\,THz, while data for Formulas D and E extend out to 1000\,THz. Because not all of the prototype filters are the same thickness, all filter transmission data were scaled to represent a 200\,\micron{} thick sample. Black vertical dashed line is at 2.5\,THz, and the red circle represents 90\% transmission.}
    \label{fig:ssolve-prototypes}
\end{figure}

SSOLVE requires good rejection of higher frequency THz light but greater than 90\% transmission at 2.5\,THz. Figure \ref{fig:ssolve-prototypes} shows a comparison of different diamond-loaded polyimide aerogel designed to meet the SSOLVE requirements. Formulas C, D, and E all meet or exceed the in-band transmission requirements for SSOLVE.  

\subsection{Microwave and Sub-Millimeter Applications}
Filters have also been designed and fabricated to meet the needs of experiments that observe in the microwave regime, like CLASS, a cosmic microwave background (CMB) polarimeter designed to observe from the ground from 33-234\,GHz in four separate frequency bands\cite{CLASS}. 

CLASS is a series of polarization-sensitive microwave telescopes in the Atacama desert in northern Chile designed to measure the CMB polarization.~\cite{CLASS_TEH_2014, CLASS} The cryogenic receivers for CLASS house custom microfabricated cryogenic detectors, transition-edge sensor (TES) bolometers, operating at a base temperature of around 50\,mK. The dilution refrigerators providing the sub-Kelvin temperature stage have limited cooling power, requiring rejection of IR radiation from warmer stages in the cryogenic chain at up to 1 part in $10^6$. Additionally, the CLASS aperture of around 40\,cm diameter requires correspondingly large filters. CLASS has been observing since 2016, producing sensitive maps of 70\% of the microwave sky in four frequency bands~\cite{CLASS_Eimer_2024, CLASS_Li_2023} and ancillary results, including measurements of the microwave brightness of Venus.\cite{CLASS_Dahal_Venus_2023} While CLASS will not employ aerogel scattering filters in the currently-fielded instruments, a spare CLASS receiver was available for testing of filters in an integrated system for potential use in future instruments. 

Filters designed for use in microwave observations are also being prototyped for the upcoming EXCLAIM instrument. EXCLAIM is a balloon-borne astrophysics mission to constrain star formation over cosmic time through measurements of integrated atomic and molecular line emission from carbon monoxide and ionized carbon from the present day to the peak of star formation approximately 10 billion years ago.~\cite{EXCLAIM_Overview_Switzer_2021, EXCLAIM_Forecast_Pullen_2023, EXCLAIM_Optics_TEH_2025} To do this, EXCLAIM uses integrated silicon spectrometers with a spectral resolving power of $R=512$ in a sub-millimeter band covering frequencies $\nu =$\,420 -- 540\,GHz coupled to a completely cryogenic ($<5$\,K) telescope. The EXCLAIM telescope sits within an open bucket helium dewar. Two aerogel scattering filters (diameters $\sim 10$\,cm) are planned to be employed in EXCLAIM to reduce IR loading on the detectors, one of which is inside an evacuated receiver cryostat, while the other is placed within the helium boil-off gas of the dewar. The open-pore structure of the aerogel lends itself naturally to operation in this environment, in which the cold helium gas can permeate and cool the aerogel scattering filter.

\begin{figure}[h!]
    \centering
    \includegraphics[width = \columnwidth]{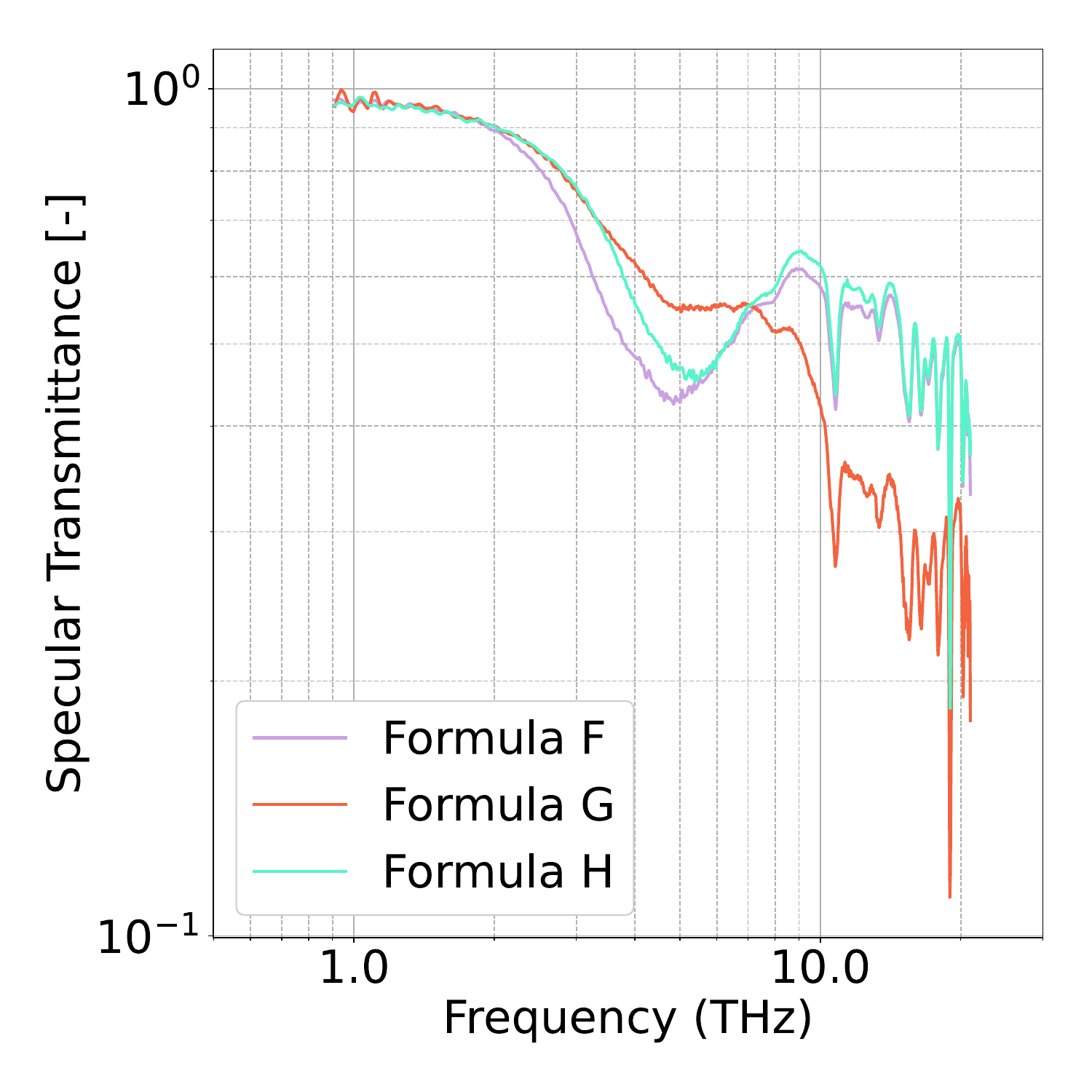}
    \caption{Transmission data for three different diamond-loaded polyimide aerogel filters for the CLASS and EXCLAIM instruments. Filter loading schema are in Table \ref{table:params}. Because not all of the prototype filters are the same thickness, all filter transmission data were scaled to represent a 1\,mm thick sample.}
    \label{fig:exclaim-prototypes}
\end{figure}

Figure \ref{fig:exclaim-prototypes} shows several different prototype filters for these two instruments. All of these filters cut off most of the infrared light above 4\,THz while yielding nearly unity transmission below 1\,THz. Formula G, likely due to the addition of 10-20\,\micron{} particles, shows lower transmission above 10\,THz than the other two formulas.

\section{Thermal modeling of filters in a receiver}
\label{section:thermal}

To predict the thermal performance of the aerogel filters in a cryogenic receiver, we simulate radiative heat transfer and heat conduction of the filter stack using finite element analysis with \textit{COMSOL Multiphysics}. Because foams and scattering filters reflect light isotropically, they are more strongly thermally-coupled to the sidewalls of a cryogenic instrument than metal-mesh filters. The stronger thermal coupling can be advantageous when combined with appropriate absorptive cryogenic coatings.

\begin{figure}[t]
    \centering
    \includegraphics[width = 0.45\textwidth]{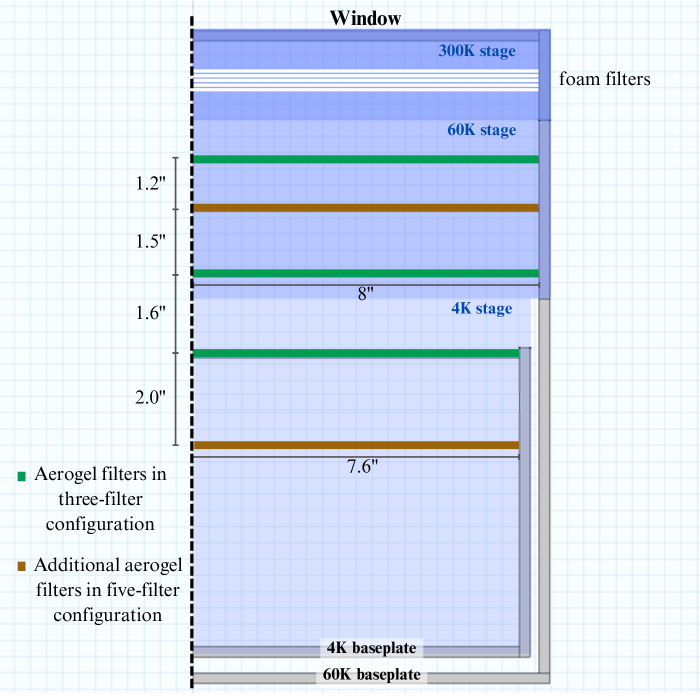}\\
    \includegraphics[width = 0.45\textwidth]{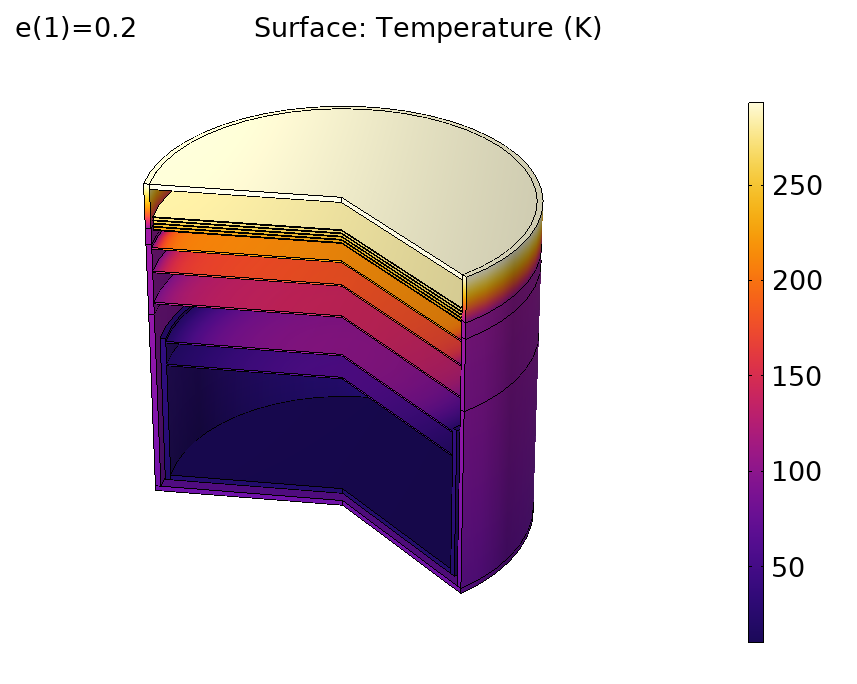}

    \caption{\textit{Top}: The geometry used for COMSOL modelling. Green lines denote the filters in a three-filter configuration, and brown lines denote the two additional filters in a five-filter configuration. White lines represent the foam filter stack behind the vacuum window. Grey lines outline the shape of the cryostat shield and baseplate. The blue dashed line represents the axis of symmetry, about which the geometry revolves and forms a 3D model of the CLASS cryogenic receiver. The window of the receiver is located at the top. \textit{Bottom}: The 3D temperature model simulated by COMSOL. The color bar shows the equilibrium temperature of the model in Kelvin. Filter emissivity is set to 0.2.}
    \label{fig:geometry}
\end{figure}

Figure \ref{fig:geometry} shows the 2D geometry of the model and the simulated equilibrium temperature under the operating conditions of a CLASS cryogenic receiver. The final 3D geometry is generated by revolving the 2D geometry about the axis of revolution, which is denoted by the blue dashed line. The CLASS cryogenic receiver employs a multistage design, where each stage is referred to by its nominal temperature, namely the 300\,K stage, 60\,K stage, and 4\,K stage. The focal plane (the coldest part of the cryogenic receiver) is located between the bottom 4\,K aerogel filter and the 4K baseplate. The two different filter configurations considered here are:
\begin{enumerate}
    \item a three-filter configuration (two filters on the 60\,K stage and one filter on the 4\,K stage)
    \item a five-filter configuration (three filters on the 60\,K stage and two filters on the 4\,K stage)
\end{enumerate}
Multiple configurations were tested because the exact emissivity and thus the exact thermal load in each scenario was unknown prior to the measurements. Both configurations include five foam filters on the 300\,K stage, denoted by the white lines under the window in Figure \ref{fig:geometry}. Note that, for simplification, the modeled 60\,K stage and 300\,K stage are connected as one geometry instead of two, and the length of the modeled receiver is also shorter than that of the actual receiver. 

\begin{figure}[h!]
    \centering
    \includegraphics[width = \columnwidth]{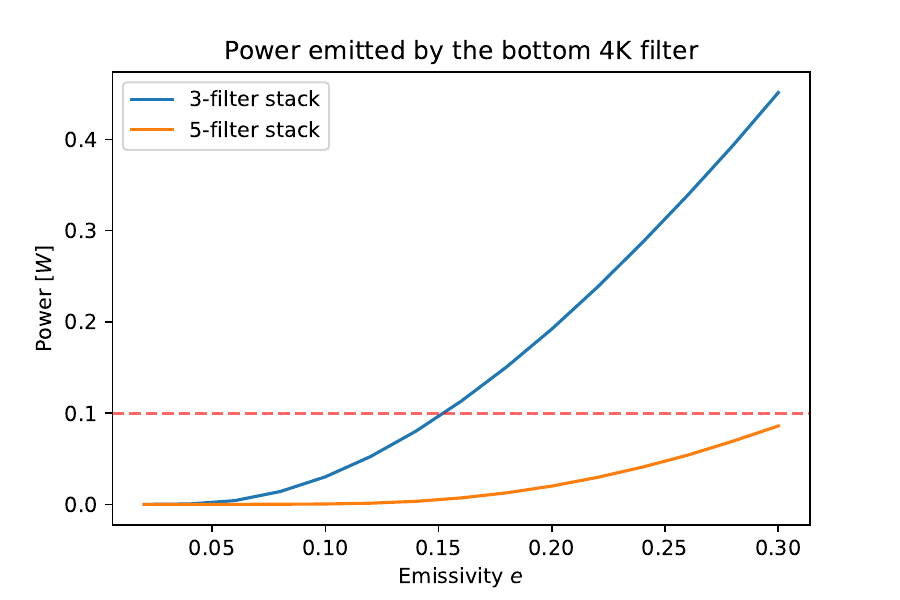}
    \caption{Power emitted by the bottom filter in the 4\,K stage as a function of filter emissivity. CLASS experiment requires a maximum load of $100$\,mW, denoted by the red dashed line.}
    \label{fig:emissivity}
\end{figure}

\begin{figure*}[t!]

    \includegraphics[width = .7\textwidth]{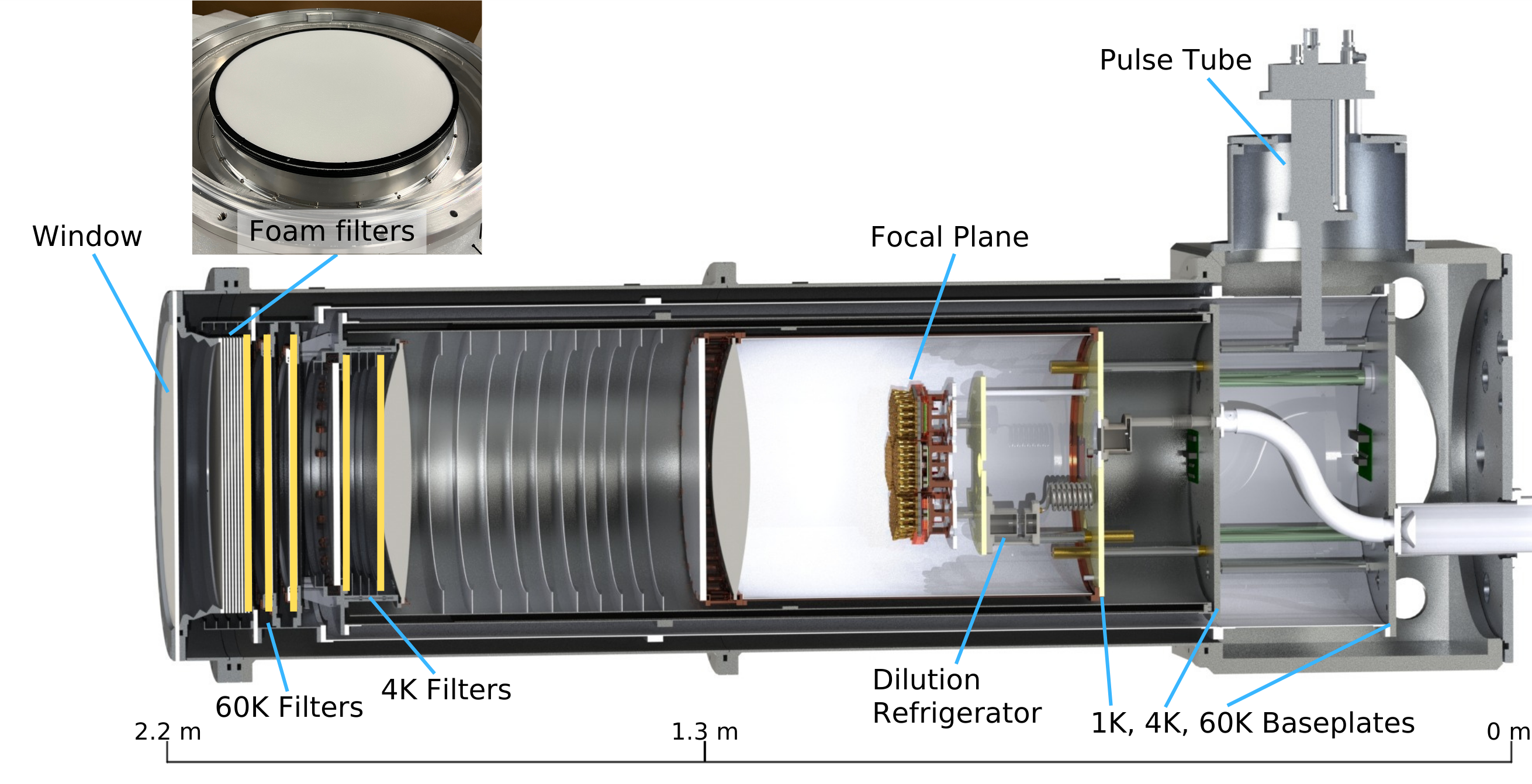}
    \includegraphics[width = .28\textwidth]{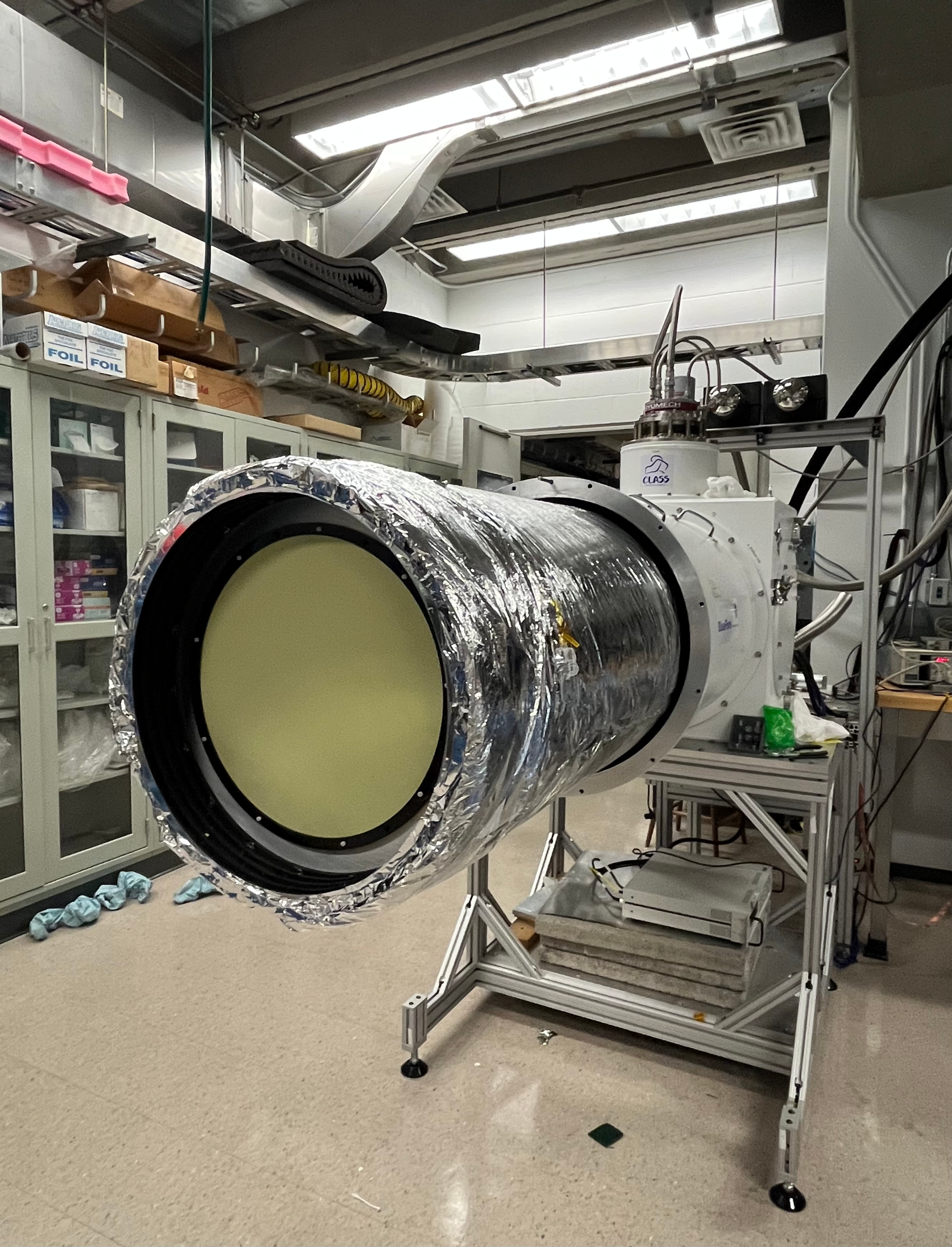}
    \caption{\textit{Left}: A diagram of CLASS 90\,GHz cryogenic receiver. Aerogel filters are represented by yellow lines. \textit{Right}: A picture showing an aerogel filter mounted on the CLASS receiver 60\,K stage. Note that all other optical components (i.e. lenses, stop etc.) in the original receiver design have been removed for the aerogel filter tests.}
    \label{fig:receiver}

\end{figure*}

These simulations consider both thermal conduction and radiation inside the region of interest. The aerogel filters are treated as uniform, perfect diffusers that transmit, radiate, and reflect uniformly in all directions. Simulations are performed iteratively for different emissivity values ranging from 0 to 0.3. Figure \ref{fig:emissivity} shows the temperature and the total emitted power of the bottom aerogel filter at the 4\,K stage as a function of emissivity. The blue line and the orange line represent the result in a three-filter configuration and five-filter configuration respectively. The red dashed horizontal line indicates the upper limit of load (0.1\,W) required by the CLASS experiment. 

We found that the simulations best match our experimental results when emissivity is between $0.15 - 0.2$, which satisfies the load requirement of CLASS if a five-filter stack configuration is adopted. An emissivity of $0.15-0.20$ roughly agrees with the estimated calculation in Figure \ref{fig:total_hemi} at 150\,THz. Broadly, the simulated numbers do not exactly agree with the estimated emissivity from Figure \ref{fig:total_hemi} because the simulation assumes perfect diffusers and the estimate is an average across all wavelengths. The simulated temperatures of the filters are quoted in brackets next to the experimental values in Table \ref{table:temp}. Knowledge of the aerogel filter emissivity will aid their implementation in future instruments. 

\begin{table*}[t!]
    \caption{Equilibrium temperature of each filter for different filter configurations. Numbers in square brackets are temperatures simulated by \textit{COMSOL}.}
    \label{table:temp}
    \resizebox{\textwidth}{!}{
    \begin{tabular}{|c|c|c|c|c|c|c|c|}
    \hline
    \hline
    \multirow{3}{*}[+1.3ex]{Filter configurations} &
    \multicolumn{5}{c|}{\multirow{2}{*}[+1.25ex]{Filter Temperature (K)}} &
    \multirowcell{2}{4\,K Baseplate \\ Temperature (K)} &
    \multirowcell{2}{4\,K Baseplate \\ Load (W)}\\
    \cline{2-6}
    
    & Top 60\,K & Middle 60\,K & Bottom 60\,K & Top 4\,K & Bottom 4\,K  &  & \\
    \hline
    Three aerogel filters only                &   238 [246] &  -  &   195 [188] &  120  [128] &  -  &  11.0 & 4.5\\
    Three aerogel filters + 300\,K foam filters &   183 [194] &   -    &  151 [145] & 99 [97]&  -  &  6.1 & 1.5\\
    Five aerogel filters only                 &  255 [253] &  210 [210] &  180 [163] &  173 [120] &  82 [80] & 9.8 & 3.8\\
    Five aerogel filters + 300\,K foam filters &   215 [217]  &  178 [180]  &  151 [141] &  149 [102]  &  73 [62] & 6.0 & 1.4\\
    \hline
    \hline
    \end{tabular}}
\end{table*}

\section{Cryogenic performance of Aerogel filters}
\label{sec:receiver_testing}
To test the performance of the aerogel filters in an actual cryogenic receiver for astronomical experiments, large-scale 1\,mm-thick aerogel filters are fabricated and tested in the CLASS 90\,GHz telescope cryogenic receiver. Figure \ref{fig:receiver} shows the diagram of CLASS 90\,GHz receiver design \cite{iuliano2018cosmology} and a picture of an aerogel filter mounted on the 60\,K stage of CLASS receiver. The CLASS cryogenic receiver uses a custom horizontal \textit{Bluefors} dilution refrigerator system with a \textit{Cryomech PT415} pulse tube cryocooler. For the tests here, the receiver is cooled with the pulse tube only, which cools the receiver to as low as $\sim$ 4\,K. Four different filter configurations have been tested: 
\begin{enumerate}[itemsep=0pt]
    \item Three aerogel filters (Two on the 60\,K stage and one on the 4\,K stage)
    \item Three aerogel filters (Two on the 60\,K stage and one on the 4\,K stage) + five foam filters on the 300\,K stage
    \item Five aerogel filters (Three on the 60\,K stage and two on the 4\,K stage)
    \item Five aerogel filters (Three on the 60\,K stage and two on the 4\,K stage) + five foam filters on the 300\,K stage
\end{enumerate}

The foam filters used are made of XPS foams that are milled into 1\,mm-thick sheets. These foam filters are installed right behind the vacuum window, as shown in Fig. \ref{fig:receiver}.

Any scattered stray light is absorbed by the blackened inner shells and baffles on both the 60\,K and 4\,K stages. The spacing and height of the baffles are set such that they are geometrically black, and all shell interiors are blackened with a mixture of Stycast epoxy and carbon lamp black.

To track the temperature of each filter during cool-down, a \textit{Cernox} Resistance Temperature Detector (RTD) is attached to the center of each filter. The equilibrium temperature of each filter and the 4\,K baseplate is summarized in Table \ref{table:temp}. The simulated temperatures are in the brackets next to the measured values. Note that in the five aerogel configurations, the measured top 4\,K filter temperature is much greater than the simulated temperature, while the others agree with the simulation quite well, which might be due to a faulty thermometer on the top 4\,K filter. Three-filter simulations yield an estimated emissivity of 0.15 and five-filter simulations yield an estimate of 0.20. 
A loading test has also been done on the CLASS 90\,GHz receiver to obtain the relationship between the thermal loading and temperature on the 60\,K and 4\,K baseplates. Because the pulse-tube cryocooler has a fixed cooling capacity, applying a known amount of Joule heating can be used to simulate infrared thermal heat load. By applying a known heat load and measuring the steady-state cryogenic temperature of the 4\,K and 60\,K baseplates, a temperature versus thermal-load relationship was determined. This is subsequently used to estimate the optical thermal load through the aerogel filters by measuring the steady-state cryogenic temperatures of the testbed.

A heater resistor was attached to the 4\,K baseplate and 60\,K baseplate respectively to heat up the baseplates with known power. The temperatures of the baseplates are measured at each given power when the cryostat is kept dark, i.e., all stages are closed up with solid metal plates. A linear function is fitted to the measured temperature-power graph. The fitted load curves are
\begin{eqnarray}
T = 1.62P + 3.68 \textrm{\,\,(4\,K baseplate)}\\
T = 1.38P + 42.46 \textrm{\,\,(60\,K baseplate)}
\end{eqnarray}
where $T$ is the temperature in Kelvin and $P$ is the power in watts. The 4\,K baseplate load inferred from the above load curve for each filter configuration is summarized in the rightmost column in Table \ref{table:temp}. Note that this inferred power is much higher than the power emitted by the bottom 4\,K filter as simulated by COMSOL. This could be because the thermal load from the window reached the baseplates through other paths, such as the gap between shields of adjacent stages, instead of going through the aerogel filter only. Another possible reason is that there are tiny air bubbles present in the aerogel filters that make them slightly more transmissive in bulk than optical measurements suggest, allowing more load to pass onto the baseplates than expected.

Despite the mismatch between the thermal load predicted by COMSOL and the cryogenic receiver tests, the filters have been shown to work cryogenically down to 4\,K and have an emissivity in the range of 0.15 -- 0.2.  

\section{Conclusion}
\label{sec:conclusion}
Diamond-loaded polyimide aerogel scattering filters are a promising emerging filter technology for use in far-infrared, sub-millimeter, and microwave astrophysics, cosmology, and planetary science missions. Prototype filters demonstrate excellent out-of-band rejection, high in-band transmission, and tunable cut-off frequencies. These filters also have low indices of refraction, negating the need for anti-reflection coatings. Unlike many COTS products, these diamond-loaded polyimide filters can be fabricated to the end user's specified thickness, instead of having to be trimmed or milled down to a desired thickness. When compared to metallic mesh filters like those manufactured by QMC Instruments Ltd., the aerogel filters have higher transmission before the filter cutoff frequency and show less high frequency light leakage. 

Filters can be manufactured with large diameters (larger than 40\,cm), while maintaining mechanical robustness. Filters have been tested in a cryogenic testbed that simulates deployment in a millimeter-wave telescope receiver. Thermal performance tests have verified filter emissivities of $ e \simeq 0.15-0.2$.

\section{Funding and Acknowledgments}
This work was supported by the National Aeronautics and Space Administration under grant numbers NNX14AB76A and NNH18ZDA001N-18-APRA18-0008, as well as the National Science Foundation Division of Astronomical Sciences for their support of this work under Grant Numbers 0959349, 1429236, 1636634, and 1654494. K.R.H. acknowledges that the material is based upon work supported by NASA under award number 80GSFC24M0006. We also gratefully acknowledge support under the Goddard Internal Research and Development (IRAD) program.

Portions of this work were presented at the SPIE conference on Astronomical Telescopes + Instrumentation in 2022, under papers titled ``Characterization of aerogel scattering filters for astronomical telescopes'' and ``Novel infrared-blocking aerogel scattering filters and their applications in astrophysical and planetary science observations.'' 

\appendix
\section{Additional Photographs and Micrographs of Foam Samples}
\label{image_appendix}
\begin{figure*}[h!]
    \centering
    \includegraphics[width = \columnwidth]{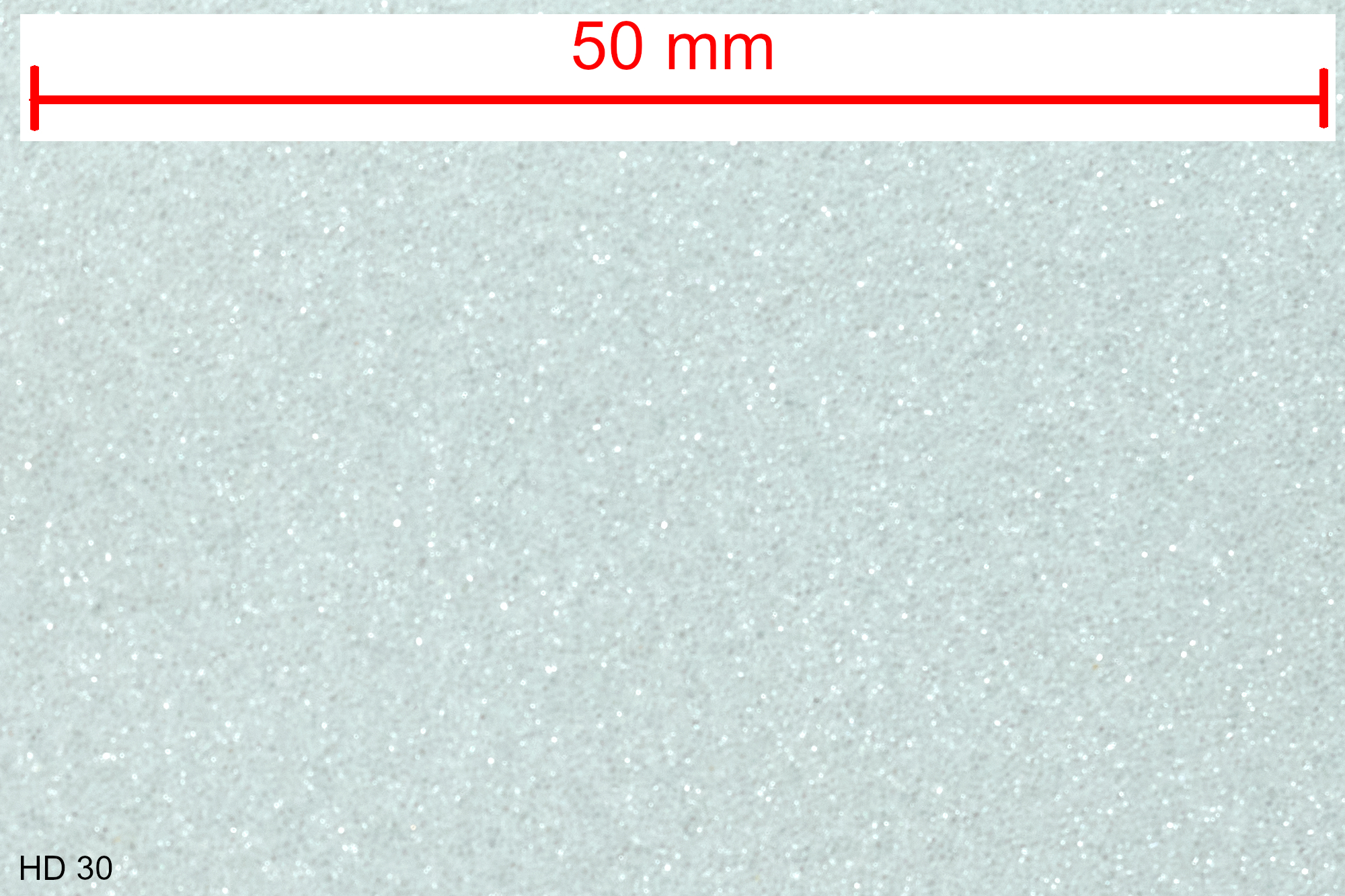}
    \includegraphics[width = \columnwidth]{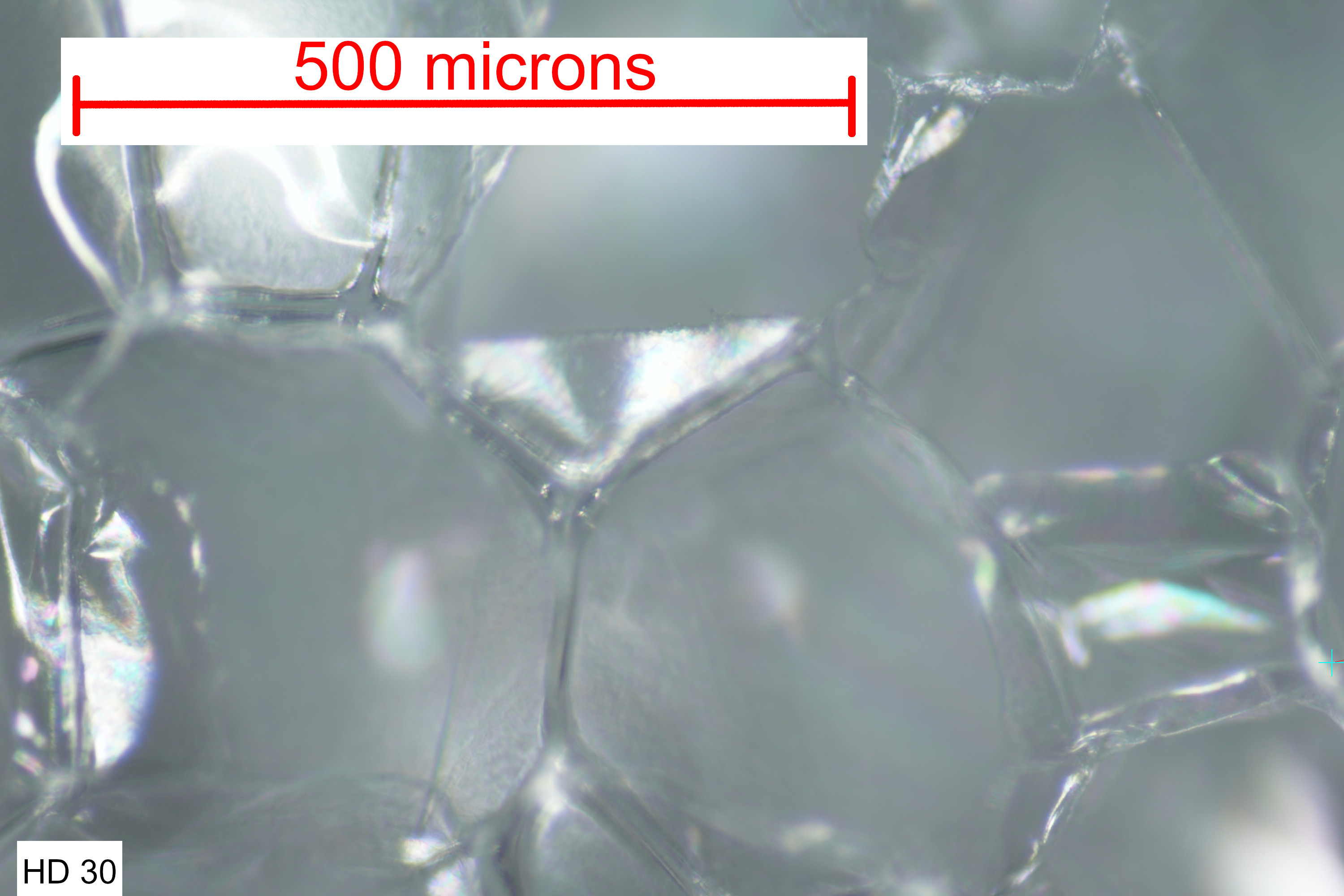}\\
    \includegraphics[width = \columnwidth]{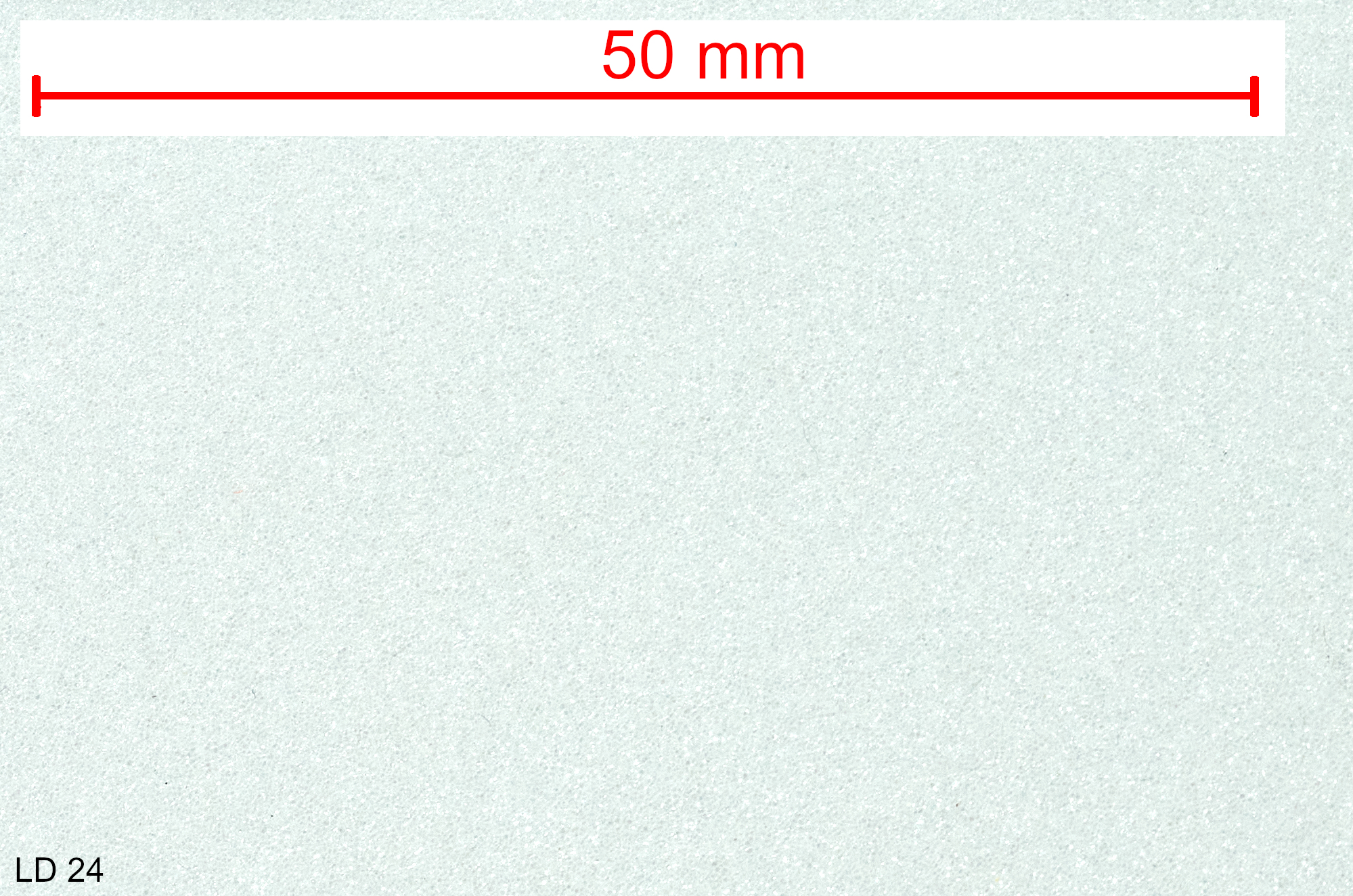}
    \includegraphics[width = \columnwidth]{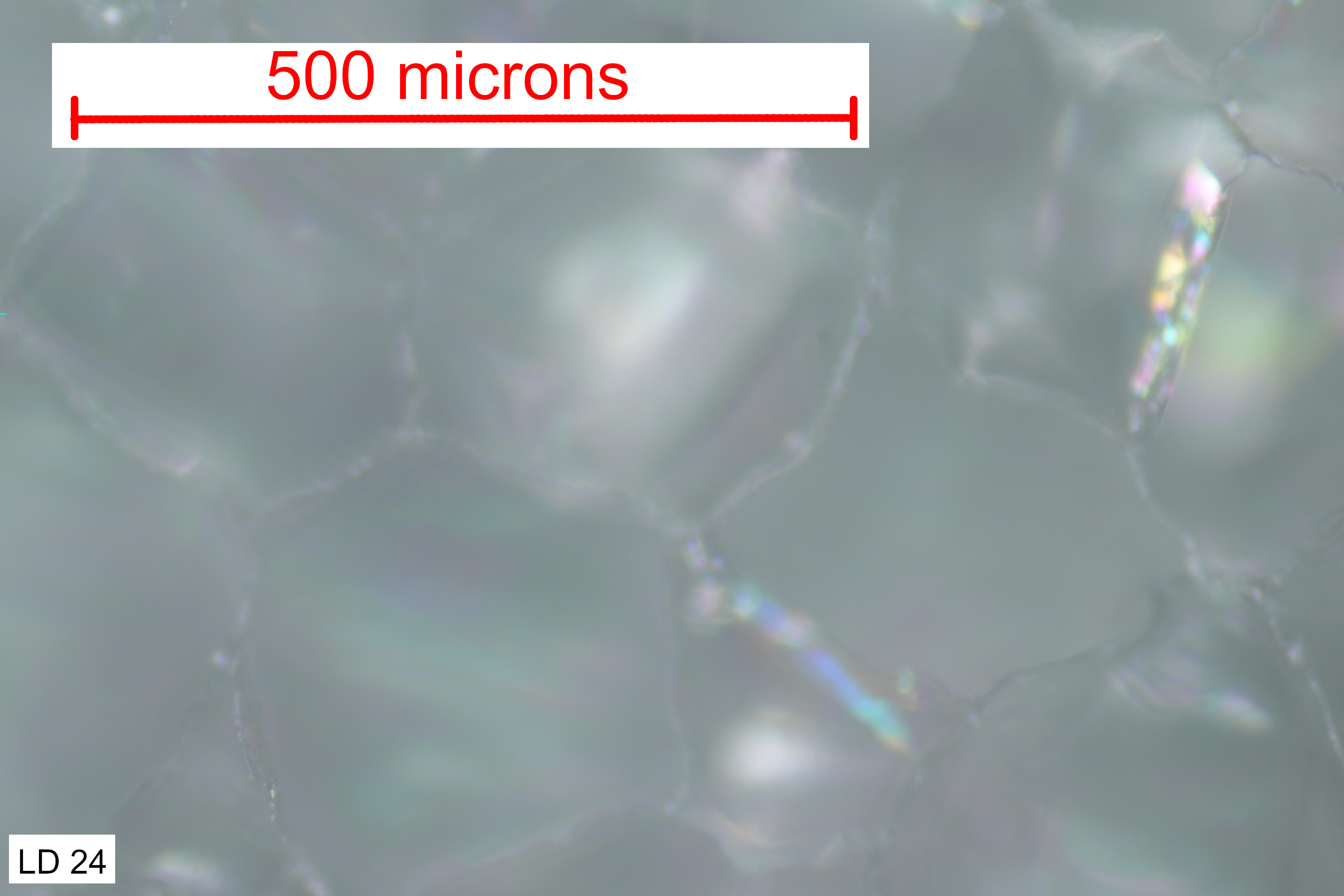}\\
    \includegraphics[width = \columnwidth]{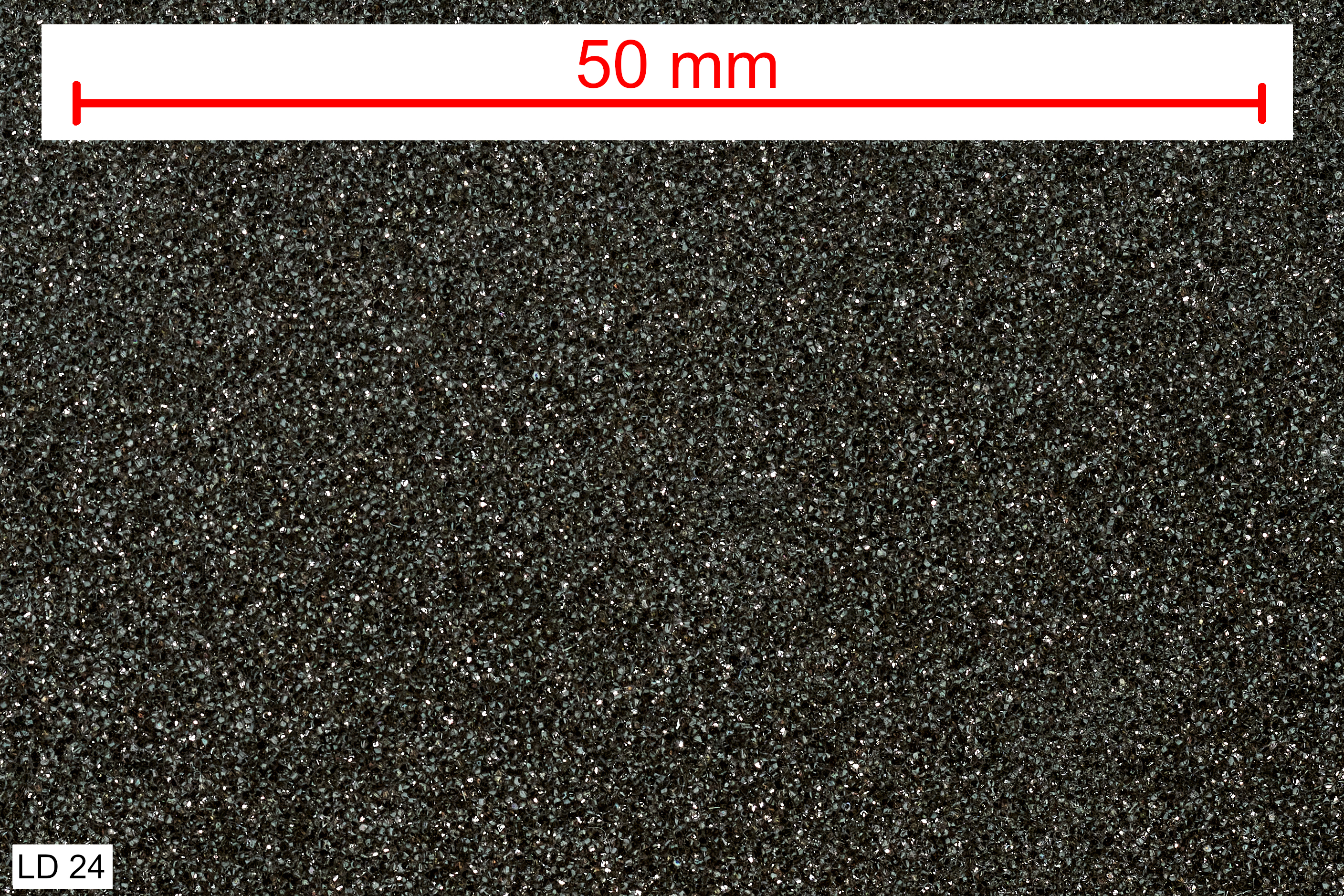}
    \includegraphics[width = \columnwidth]{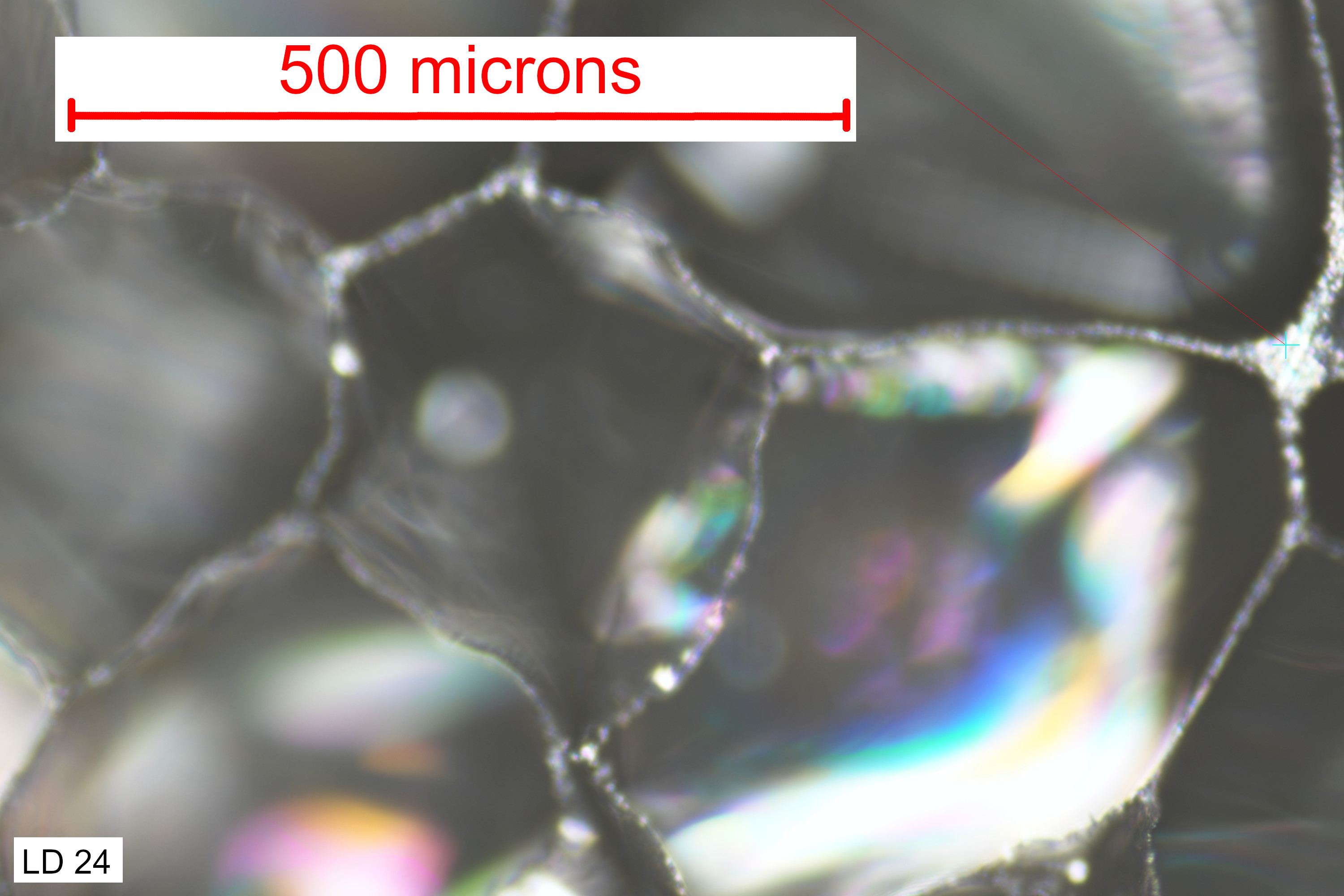}
    \caption{Macroscopic (left) and microscopic (right) images of samples of Zotefoam. Top row is white Zotefoam HD30, middle row is white Zotefoam LD24, and bottom row is black Zotefoam LD24}
    \label{fig:whiteHD30}
\end{figure*}

\clearpage

\bibliography{main}

\end{document}